\documentclass{emulateapj}
\usepackage{rotating}
\usepackage{amsmath}
\usepackage{natbib}
\usepackage{threeparttable}
\pdfoutput=1

\newcommand{\be}{\begin{equation}}
\newcommand{\ee}{\end{equation}}
\newcommand{\ba}{\begin{eqnarray}}
\newcommand{\ea}{\end{eqnarray}}

\def\eg{{e.g.,}}
\def\ie{{\it i.e. }}
\def\hp{{H$^{+}$}}
\def\hep{{He$^{+}$}}
\def\hetwo{{He$^{2+}$}}
\def\cp{{C$^{+}$}}
\def\ctwo{{C$^{2+}$}}
\def\cthree{{C$^{3+}$}}
\def\cfour{{C$^{4+}$}}
\def\ofive{{O$^{5+}$}}
\def\oone{{O$^{+}$}}
\def\otwo{{O$^{2+}$}}
\def\othree{{O$^{3+}$}}
\def\ofour{{O$^{4+}$}}
\def\ofive{{O$^{5+}$}}
\def\osix{{O$^{6+}$}}
\def\oseven{{O$^{7+}$}}
\def\mgp{{Mg$^{+}$}}
\def\mgtwo{{Mg$^{2+}$}}
\def\nap{{Na$^{+}$}}
\def\elec{{e$^{-}$}}
\begin{document}
\title{Atomic Chemistry in Turbulent Astrophysical Media I: Effect of Atomic Cooling}
\author{William J. Gray\altaffilmark{1}, Evan Scannapieco\altaffilmark{2}, \& Daniel Kasen\altaffilmark{3,4}}

\altaffiltext{1}{Lawrence Livermore National Laboratory, P.O. Box 808, L-038, Livermore, CA 94550, USA}
\altaffiltext{2}{School of Earth and Space Exploration, Arizona State University, P.O. Box 871404, Tempe, AZ 85287-1494, USA}
\altaffiltext{3}{Lawrence Berkeley National Laboratory, Berkeley, CA 94720, USA}
\altaffiltext{4}{Department of Physics, University of California, Berkeley, CA 94720, USA}
\slugcomment{{\sc Accepted to ApJ}}

\begin{abstract}

We carry out direct numerical simulations of turbulent astrophysical media that explicitly track ionizations, recombinations, and species-by-species radiative cooling. The simulations assume solar composition and follows the evolution of hydrogen, helium, carbon, oxygen, sodium, and magnesium, but they do not include the presence of an ionizing background.   In this case, the medium reaches a global steady state that is purely a function of the one-dimensional turbulent velocity dispersion, $\sigma_{\rm 1D},$ and the product of the mean density and the driving scale of turbulence, $n L.$   Our simulations span a grid of models with $\sigma_{\rm 1D}$ ranging from 6 to 58 km s$^{-1}$ and $n L$ ranging from 10$^{16}$ to 10$^{20}$ cm$^{-2},$ which correspond to turbulent Mach numbers from $M=0.2$ to 10.6.   The species abundances are well described by single-temperature estimates whenever $M$ is small,  but local equilibrium models can not accurately predict the global equilibrium abundances  when $M \gtrsim 1.$    To allow future studies to account for nonequilibrium effects in turbulent media, we gather our results into a series of tables, which we will extend in the future to encompass a wider range of elements, compositions, and ionizing processes.

\end{abstract}

\section{Introduction}

Turbulence is ubiquitous in astrophysics, where the  Reynolds number (Re), the ratio of the inertial forces to the viscous forces,  is often orders of magnitudes higher than found on the Earth.   In the intergalactic medium, for example, Re typically exceeds $10^5,$ and in the warm ionized interstellar medium Re $\gtrsim 10^7$  \citep{Braginskii1958, Spitzer1962}, whereas the transition from laminar to turbulent flow occurs at Re $\approx 3 \times 10^3$  \citep[\eg][]{Orszag1980}. In addition, in many astrophysical regimes, rapid cooling causes turbulent velocities to exceed the sound speed, and within such supersonic turbulence, random motions can  compress a fraction of the material to very high densities \citep[\eg][]{Padoan1997, MacLow2004, Federrath2008}, producing a complex, multi-phase medium.

Further complicating the picture is the fact that the recombination and collisional ionization times for many species are long with respect to the ``eddy turn-over time" on which existing turbulent motions decay and new turbulent motions are added.  Thus, the conditions experienced by a parcel of gas may change before equilibrium can be reached, such that the ionization structure of the medium will depend not only on  the small scale density and temperature distribution, but on the velocity distribution as well. For these reasons,  the turbulent structure of gas can have a significant impact on line emission and absorption diagnostics.  Most interpretations of observed spectra, however, do not take into account these multi-phase and non-equilibrium effects, and full galaxy simulations can not resolve the relevant small-scale structures to include them.

% in which viscosity is caused by the transport of momentum by ions  \citep{Braginskii1958, Spitzer1962},  
%In an ionized medium, in which viscosity is caused by the transport of momentum by ions,  ${\rm Re} = 8.5 \times 10^5 (v/{\rm km \, s^{-1}}) (L/{\rm parsec}) (n/{\rm cm}^{-3}) (T/10^4 K)^{2.5}$ where $v$ and $L$  are the velocity and length scale on which the flow is driven, and  $n$ and $T$ are the temperature and density of the medium \citep{Braginskii1958, Spitzer1962}. In the warm ionized interstellar medium, where $n \approx 1$ cm$^{-3}$ and $T \approx 10^4$K,  Re can regularly exceed $10^7$, and in the larger scale ($L\approx 100$ pc) but lower density ($n \approx 10^{-4} {\rm cm}^{-3})$ intergalactic medium, it can exceed $10^5$, whereas the transition from laminar to turbulent flow occurs at Re $\approx 3 \times 10^3$.
%These Re values can be boosted even further by the presence of magnetic fields, which cause ions to travel along helical paths with gyro-radii many orders of magnitude smaller than the collisional mean free path (Spitzer 1962).  This leads to a strongly-anisotropic viscosity that can increase the effective Reynolds number  by a factor of $1-10,$ depending on the field configuration \citep[\eg][]{Narayan2001, Reynolds2005}.

Furthermore, there are several reasons to expect that nonequilibrium effects are important in interpreting current observations. For example, emission line studies of high star-formation rate, ultraluminous infrared galaxies (ULIRGs) have progressed to the point that a number of sensitive line diagnostics can now be considered.  Recently, \cite{Soto2012} were able to measure emission lines from material in and around 39 ULIRGs, using the measurements to construct line ratios that are relatively insensitive to the presence of dust, but highly constraining of the sources of photoionization \citep[\eg][]{Kewley2006, Allen2008}. They found that material up to $\approx 10$ kpc from the galaxy centers was primarily heated by shocks rather than photoionization, as would be expected for strongly turbulent media.

At larger distances, \cite{Werk2014} used the {\em Cosmic Origins Spectrograph} \citep{Green2012}  to measure low and intermediate ionization state ions in the circumgalactic medium (CGM) within $\approx 100$ kpc of low-redshift galaxies \citep[see also, ][]{Tumlinson2013,Werk2013,Peeples2014}. To model these absorbers they adopted several important assumptions, namely: (i) that the ions were co-spatial and arose from a single phase; (ii) that the medium was in ionization equilibrium; and (iii) that the medium was primarily photoionized. Surprisingly, to match the observations with such assumptions, the models had to adopt large ionization parameters, defined as the ratio of ionizing photon density to the hydrogen density. Given the observed range of metagalactic and host galaxy fluxes, these ratios corresponded to densities and pressures over two orders of magnitude lower than expected from hydrostatic balance. However, given the long recombination and cooling times in the diffuse CGM, even moderate energy input from decaying turbulence may be sufficient to substantially change this picture.

It is with these  issues in mind that we have carried out a comprehensive numerical survey of the atomic structure and observational properties of turbulent astrophysical media. While a large number of high-resolution studies of turbulence exist in the literature, the majority of these are incompressible simulations carried out in the context of fluid-dynamics research  \citep[\eg][]{Vincent1991, Moin1998, Ishihara2009}, and compressible, isothermal simulations carried out in the context of interstellar medium research \citep[\eg][]{Kritsuk2007, Federrath2008, Schmidt2009, Pan2010, Sur2014}.  \cite{Walch2011} carried out simulations of both continuously-driven and decaying turbulence in solar metallicity and 0.001 solar metallicity material in a medium with a one-dimensional turbulent velocity dispersion, $\sigma_{\rm 1D} \approx 30$ in a 500 parsec cubical box with a mean hydrogen density of 1 cm$^{-3},$ using the chemical network of \cite{Glover2007} to track the nonequilibrium chemistry associated with hydrogen and helium \citep[see also][]{Walch2014}. \cite{Saury2014}, on the other hand, studied the structure of thermally bistable continuously-driven turbulent  gas, using a cooling rate given as  $n^2 \Lambda(T)$ and a heating rate given as $n \Gamma(T),$ where $n$ was the local number density 
and $\Lambda$ and $\Gamma$ were global functions accounting for  atomic cooling, recombination on interstellar grains, and photo-electric heating of small dust grains \citep{Wolfire2003}. 

Here we carry out exact calculations of atomically-cooled astrophysical media that explicitly track continuously-driven turbulent motions, radiative cooling by atomic species,  and the nonequilibrium ionization and recombination of several elements, for a grid of solar-metallicity models with $\sigma_{\rm 1D}$ ranging from 6 to 58 km s$^{-1}$ and the product of the mean density and the turbulent driving scale ranging from 10$^{16}$ to 10$^{20}$ cm$^{-2}.$  In this first paper in this series, we track the  nonequilbrium ionization structure of hydrogen, helium, carbon, oxygen, sodium, and magnesium, focusing on the limit in which ionization is purely collisional. Taking advantage of the scaling properties of this case, we are able to fully span the relevant range of conditions experienced in atomically cooled astrophysical plasmas: cataloging their physical properties for comparison with more idealized simulations and tabulating their species mass fractions and Doppler parameters for use in the interpretation of future theoretical and observational studies.    
  
The structure of the paper is as follows. In \S2 we describe our atomic chemistry and cooling routines and their associated tests. In \S3 we present our simulation setup and initial conditions, and in \S4 we describe our results, taking particular note of their probability density functions and the effect of resolution and thermal conduction. Concluding remarks are given in \S5.

\section{Numerical Method}

All simulations were performed with FLASH version 4.0.1 \citep{Fryxell2000}, a publicly-available hydrodynamics code. To ensure the stability of the code as turbulence develops, we employ a hybrid Riemann solver which uses both an extremely accurate but somewhat fragile Harten-Lax-van Leer-Contact (HLLC) solver \citep[\eg][]{Toro1994, Toro1999} and a more robust, but more diffusive Harten Lax and van Leer (HLL) solver \citep{Einfeldt1991}. The HLLC solver is a modification to the HLL solver that includes the missing shear and contact waves, and it produces solutions that most accurately capture contact discontinuities. However, in regions with strong shocks or rarefactions, the HLLC solver can fail, and in such situations, we switch to the positivity-preserving HLL solver.  Magnetohydrodynamic effects were not included in this study.

In order to accurately determine the atomic properties of turbulent media, we also added two new capabilities to the code: a non-equilibrium ionization package that tracks the ionization state of several atomic species, and a cooling routine that takes into account the cooling from each of these ionized states individually. In this section we describe our numerical implementation of each of these new capabilities. 

\subsection{Atomic Ionization}
 
Our ionization network tracks the impact of 36 separate reactions acting on 24 species and 6 elements: hydrogen (H, \hp), helium (He, \hep, \hetwo), carbon (C-\cfour), oxygen (O-\oseven), sodium (Na, \nap), magnesium (Mg, \mgp, \mgtwo), and free electrons (\elec). For each species, we track several reactions, including radiative recombinations, dielectronic recombinations, and ionization due to electron impacts. A summary of the reactions considered as well as their source is given in Table~\ref{speclist}.

Our implementation of this network follows the overall approach we have adopted in previous studies \citep{Gray2010, Scannapieco2012, Gray2013}. Each species is labeled by an index $i$, such that species $i$ has Z$_i$ protons, A$_i$ nucleons, and a mass density of $\rho_i$. We then define a mass fraction of species $i$ as X$_i$ $\equiv$ $\rho_i/\rho$, where $\rho = \sum_i \rho_i$, and define the molar mass fraction as Y$_i$ $\equiv$ X$_i$/A$_i$. For each species, a continuity equation for the molar mass fraction is then given as $\dot{Y}_i = R_i$, where $R_i$ is the total reaction rate due to all reactions.

Because of the complex ways that reaction rates depend on temperature and because of the large range of possible abundances, the resulting rate equations are often `stiff,' $\ie$, the change in timescales can be very different from one species to another. This requires implicit or semi-implicit methods to track all the relevant reactions throughout our simulations.  To handle this, we use a variable-order Bader-Deuflhard method \citep[\eg][]{Bader1983} which is well suited to problems with dimension $>10$ \citep[\eg][]{Press1992,Bovino2013}. The network presented here also has the advantage of being sparse. That is, the Jacobian, defined as $J_{i,j} \equiv \partial{ \dot{Y_i} } / \partial{Y_j}$, is mostly comprised of zeros, with nonzero values falling on or near the diagonal. This allows us to use the MA28 sparse linear algebra package included with FLASH \citep{Duff1986} to compute the matrix inverses required by the Bader-Deuflhard scheme, leading to a very fast and efficient implementation. In general, the chemistry package runs slightly faster than the hydrodynamics.
 
As the species evolve over a given time step, the temperature can change as the internal energy changes from ionizations and recombinations. Since the reaction rates are strong functions of temperature, the network can become unstable if too large a time step is used, especially in regions of strong cooling. To ensure testability  but allow the simulation to evolve at the hydrodynamic time step, we subcycle the rate equations within each cell on a network time step,
defined as:
\be
t_{\rm net} \equiv \rm{ min} \left(\alpha \frac{Y_i + 0.1 Y_H^+ } { \dot{Y_i} } \right),
\ee
where $\alpha$ is a constant that controls the maximum abundance change allowed, which we default to 0.1. Using the current species abundances, the instantaneous change in abundances $\dot Y_i$ is computed using the analytic ordinary differential equations at the current temperature. Note that a small fraction of $Y_H^+$ is added as a buffer to prevent species with very small, but rapidly changing, abundances from causing prohibitively small time steps and excessive subcycling. In addition to this network subcycling, the Bader-Deuflhard method includes its own internal subcycling and time step controls.

At the beginning of each cycle, $t_{\rm net}$ is compared to the hydrodynamic time step.  If the hydrodynamic time step is smaller, then no subcycling is done and the hydrodynamic time step is used to update the molar mass fractions. On the other hand, if the network time step is smaller, then the molar mass fractions are updated using the $t_{\rm net}$, which is then recalculated and compared to the remaining hydrodynamic time step. This cycle continues until the network has advanced for the full hydrodynamic time step. 

To test the implementation of our network, we emulated a series of equilibrium models by fixing the density and temperature in the simulation  and running each case until the species reached equilibrium. We ran a series of seven such models that spanned the temperature range between 10$^4$ and 10$^7$ K, at a fixed density of 2.0$\times$10$^{-20}$ g cm$^{-3}$. All species were assumed to be neutral at the start of each simulation, and the final abundances were  compared to equilibrium  models from Cloudy (version 10.01) \citep{Ferland1998}, using the ``coronal equilibrium'' command that enforces only collisional ionizations. 

\begin{figure*}
\begin{center}
\includegraphics[trim=0.0mm 0.0mm 0.0mm 0.0mm, clip, scale=0.85]{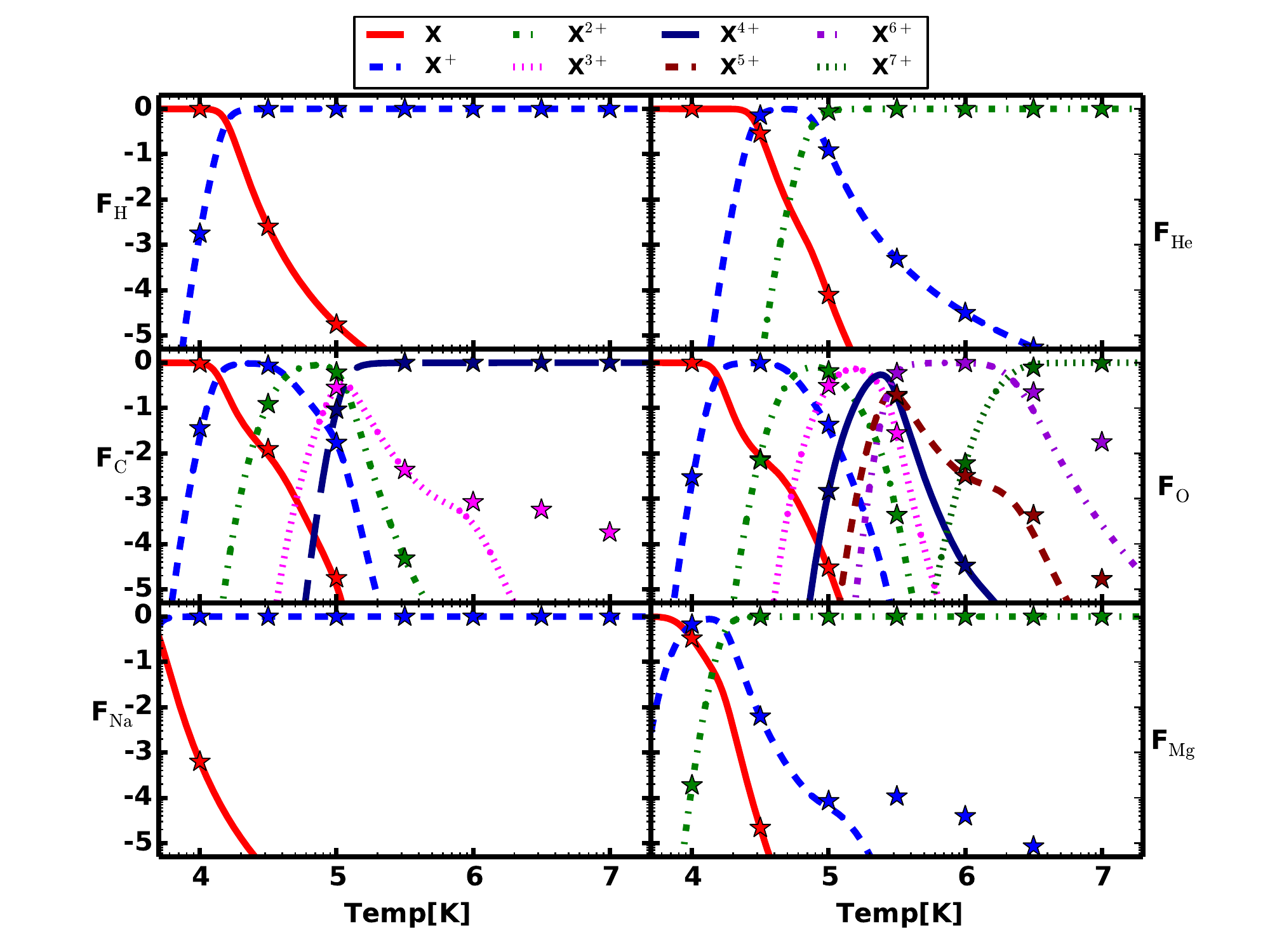}
\caption{Comparison of the species abundances between Cloudy and FLASH. Each panel shows the results for a different set of atomic species. The lines corresponds to the Cloudy results, while the points are from FLASH. We plot the equilibrium temperature along the $x$-axis and the fractional abundance of each species, \ie $F_i = n_i/n_s$ where $n_s$ is the elemental abundance, for each species as the $y$-axis. A universal legend is given at the top of the figure and the same ionization state is given by the same color and line style in each panel. {\it Top Left:} Hydrogen. {\it Top Right:} Helium. {\it Middle Left:} Carbon. {\it Middle Right:} Oxygen. {\it Bottom Left:} Sodium. {\it Bottom Right:} Magnesium. }
\label{chemtest}
\end{center}
\end{figure*}

We found that we matched the results from Cloudy very well over a wide range in temperature for all species. However, we did not match precisely for certain higher ionization states, as seen, for example, in the middle panel of Fig.~\ref{chemtest}, but this is to be expected since our network does not follow every ionization state for some elements. We did find excellent agreement for those elements for which the ionization state changes by orders of magnitude with a modest increase in temperature. For the purposes of comparison, we summed together the higher ionization states that we do not follow and assigned their abundance to the highest state that we do follow. For example, we track the abundance of C$^{4+}$ and not the two higher states. Therefore, the (dashed) navy line in the middle left panel of Fig.~\ref{chemtest} represents the summation of \cfour, C$^{5+}$, and C$^{6+}$.  

\subsection{Cooling}

The second capability we implemented is individual cooling rates associated with each ion in the chemical network. To do this, we followed the procedure presented in \cite{Gnat2012} who compiled the ion-by-ion cooling efficiencies for several atomic species for between 10$^{4}$-10$^{8}$ K. These efficiencies include all the cooling processes considered in Cloudy, as described in \cite{Osterbrock2006}. This includes collisional excitations followed by line emission, recombinations with ions, collisional ionizations, and thermal bremsstrahlung. Here we briefly outline this procedure, which we used to extend their work down to 5$\times$10$^{3}$ K for the ions in our network. Again, making use of Cloudy, we construct a grid of models that span 5$\times$10$^{3}$ to 10$^{8}$ K in temperature for each ion $i$ of element $E$. Each model is then composed of only hydrogen, electrons, and the ion under consideration. To ensure cooling from hydrogen is suppressed, the hydrogen density is set to 10$^{10}$ times smaller than the ion density. Following \cite{Gnat2012}, the elemental ion and electron density is set to $n_{E_i}$ = $n_{e^-}$ = 1.0 cm$^{-3}$ and $n_{H}$ = 10$^{-10}$ cm$^{-3}$. This produced a cooling rate for each ion as a function of temperature, and a comparison of these rates to those of \cite{Gnat2012} show identical results where the temperature range overlapped.
 
For those species in which we do not follow every ionization state in our simulation individually (such as C$^4+$, which captures the joint abundance of C$^{4+}$, C$^{5+}$, and C$^{6+}$) we formed a composite cooling curve as
\be
\Lambda(T) = \frac{ \sum_i n_i(T) \Lambda_i(T) } { \sum_i n_i(T)},
\ee
where $\Lambda_i(T)$ is the cooling rate of ion $i$ and $n_i(T)$ is the relative abundance of ion $i$ in equilibrium at a temperature $T$. This composite is then used as the cooling rate for the highest ionization state followed.
Similarly, we included cooling from nitrogen, neon, silicon, sulphur, calcium, and iron. Again, we assumed that the relative ionization abundances were determined by the collisional ionization equilibrium. This final cooling curve has the form 
\be
\Lambda_{\rm other}(T) = \sum_j \frac{ \sum_i n_{j,i} (T) \Lambda_{j,i} (T)}{ \sum_i n_{j,i} (T)}, 
\ee
where $j$ loops over nitrogen, neon, silicon, sulphur, calcium, and iron.

With the addition of cooling, an important timescale is created, namely one that relates the total internal energy to energy loss rate per time as
\be
t_{\rm cool} = \frac{ \alpha_c E_i}{\dot{E_i} },
\ee
where $E_i$ is the internal energy, $\dot{E_i}$ is the energy loss rate, and $\alpha_c$ is a constant between 0 and 1. As is the case with reaction rates, cooling rates are strong functions of temperature and species abundances, both of which can change drastically over a single chemistry time step. Therefore, we employ a method of subcycling the cooling on a cooling timescale within the chemistry routine. At the beginning of each cooling cycle, we calculate the cooling timescale using a given $\alpha_c$(=0.1). If this is shorter than the chemistry time step, we cycle over the cooling timescale, recalculating the temperature, cooling rate, and timescale at the end of each cycle. This proceeds until we have reached the chemistry time step. 

Since we only consider two body interactions, the cooling time should scale linearly with density.  To test this scaling, along with the overall dependence of our routines on temperature, we ran several tests in which the medium was initially completely ionized and the initial temperature was set to 1$\times$10$^{6}$ K, each of which had a  different density between 5.0$\times$10$^{-27}$ and 1.0$\times$10$^{-25}$  g cm$^{-3}$.   
%The results of these runs are shown in Fig.~\ref{cooltest}, which shows the temperature evolution versus time. 
These tests recovered the expected scaling with density, and also confirmed that cooling  was always efficient at high temperatures and much less efficient at  $\approx$10$^{4}$ K, when most elements become neutral.  Finally, we performed additional tests to ensure that the interpolation was smooth and always reproduced the table values under the correct conditions.

\begin{table}
\caption{List of the reactions considered and their source.}
\begin{threeparttable}
\centering
\begin{tabular}{lll}
\hline
Number & Reaction & Source \\
\hline
\hline
0001   & \hep    + \elec $\rightarrow$ He                & 1 \\
0002   & \cp     + \elec $\rightarrow$ C                 & 5 \\
0003   & \ctwo   + \elec $\rightarrow$ \cp               & 4 \\
0004   & \cthree + \elec $\rightarrow$ \ctwo             & 3 \\
0005   & \cfour  + \elec $\rightarrow$ \cthree           & 2 \\
0006   & \oone   + \elec $\rightarrow$ O		         & 7 \\
0007   & \otwo   + \elec $\rightarrow$ \oone		     & 6 \\
0008   & \othree + \elec $\rightarrow$ \otwo		     & 5 \\
0009   & \ofour  + \elec $\rightarrow$ \othree           & 4 \\
0010   & \ofive  + \elec $\rightarrow$ \ofour            & 3 \\
0011   & \osix   + \elec $\rightarrow$ \ofive            & 2 \\
0012   & \oseven + \elec $\rightarrow$ \osix             & 1 \\
0013   & \mgp    + \elec $\rightarrow$ Mg                & 8 \\
0014   & \mgtwo  + \elec $\rightarrow$ \mgp              & 9 \\
0015   & \nap    + \elec $\rightarrow$ Na                & 8 \\
0016   & \hp     + \elec $\rightarrow$ H                 & 12 \\
0017   & H       + \elec $\rightarrow$ \hp     + 2 \elec & 11 \\
0018   & He      + \elec $\rightarrow$ \hep    + 2 \elec & 11 \\
0019   & \hep    + \elec $\rightarrow$ \hetwo  + 2 \elec & 11 \\
0020   & C       + \elec $\rightarrow$ \cp     + 2 \elec & 11 \\
0021   & \cp     + \elec $\rightarrow$ \ctwo   + 2 \elec & 11 \\
0022   & \ctwo   + \elec $\rightarrow$ \cthree + 2 \elec & 11 \\
0023   & \cthree + \elec $\rightarrow$ \cfour  + 2 \elec & 11 \\
0024   & O       + \elec $\rightarrow$ \oone   + 2 \elec & 11 \\
0025   & \oone   + \elec $\rightarrow$ \otwo   + 2 \elec & 11 \\
0026   & \otwo   + \elec $\rightarrow$ \othree + 2 \elec & 11 \\
0027   & \othree + \elec $\rightarrow$ \ofour  + 2 \elec & 11 \\
0028   & \ofour  + \elec $\rightarrow$ \ofive  + 2 \elec & 11 \\
0029   & \ofive  + \elec $\rightarrow$ \osix   + 2 \elec & 11 \\
0030   & \osix   + \elec $\rightarrow$ \oseven + 2 \elec & 11 \\
0031   & Mg      + \elec $\rightarrow$ \mgp    + 2 \elec & 11 \\
0032   & \mgp    + \elec $\rightarrow$ \mgtwo  + 2 \elec & 11 \\
0033   & Na      + \elec $\rightarrow$ \nap    + 2 \elec & 11 \\
0034   & \hetwo  + \elec $\rightarrow$ \hep              & 12 \\
0035   & \hep    + H     $\rightarrow$ He      + \hp     & 12 \\
0036   & He      + \hp   $\rightarrow$ \hep    + H       & 12 \\
\label{speclist}
\end{tabular}
\begin{tablenotes}
\item[] \textbf{Notes.} Radiative Recombination rates taken from \cite{Badnell2006RR}. (1)\cite{Badnell2006H}, (2)\cite{Bautista2007}, (3)\cite{Colgan2004}, (4)\cite{Colgan2003}, (5)\cite{Altun2004},(6)\cite{Zat2004a}, (7)\cite{Mitnik2004}, (8)\cite{Zat2004b}, (9)\cite{Altun2006}, (10)\cite{Verner1996}, (11)\cite{Voronov1997}, and (12)\cite{Glover2008}.
\end{tablenotes}
\end{threeparttable}
\end{table}

\section{Model Framework \& Initial Conditions}

With these procedures in place, we carried out a suite of simulations of turbulent media under a wide variety of conditions. To drive turbulence, we made use of an artificial forcing term $\bf{F},$ incorporated into the momentum equation as
\be
\frac{\partial{\rho \bf{v}}}{\partial{t}} + \nabla (\rho \bf{v}\bf{v} ) + \nabla P = \rho \bf{F},
\ee
where $\rho$ is the density, $P$ is the pressure, and $\bf{v}$ is the velocity. The forcing term was modeled as a stochastic Ornstein-Uhlenbeck process  \citep{Uhlenbeck1930, Schmidt2009, Federrath2010, Pan2010} with a user-defined forcing correlation time $t_f$. For all the simulations presented here, turbulence was driven solely though solenoidal modes (\ie $\nabla \cdot {\bf F} = 0$) in the range of wavenumbers 1 $\le  L_{\rm box} |{\bf k}|/2 \pi \le$ 3, such that the average forcing wavenumber was $k_f^{-1}   \simeq 2 L_{\rm box}/2 \pi,$ with $L_{\rm box}$ the size of our turbulent box, which was fixed at 100 parsecs on a side.  This turbulence was always continuously driven throughout the simulation runtime.

All our simulations were performed using the multispecies extension for the ideal gas equation of state, which calculates the important thermodynamic quantities based on the properties of the included species \citep{Colella1985}. In particular, the average atomic mass can change dramatically as the gas either recombines or becomes ionized. While FLASH has an adaptive mesh capability, this is capability is not useful in a case such as ours in which structures of interest are distributed throughout the full simulation volume. Thus all simulations were performed using a cubic uniform grid with periodic boundary conditions. 

The material is assumed to have solar metallicity, and each run is defined by an initial uniform density and the strength of turbulent forcing. The ionization state of each ion is initially set to be consistent with a 10$^5$ K gas in collisional ionization equilibrium, except for the lowest velocity dispersion runs where all species were assumed to be neutral with an initial temperature of 10$^4$ K. For most cases, the eddy turnover timescale was much shorter than the timescale for the chemistry to come to equilibrium. To prevent long run times, we implemented a method of accelerating the chemistry such that it reached the steady state solution in a much shorter number of cycles. Once the turbulence had reached a steady state, normally after a few eddy turnover times, we carried out a `kick cycle' over which the chemistry was evolved for a much longer time than either the local hydrodynamic time step or the eddy turnover timescale. The result of this procedure was to force each cell to nearly reach collisional ionization equilibrium. Each model was then run normally until it reached a global steady state in terms of both the hydrodynamic variables and the chemical abundances. 

In reactions that involve free electrons recombining with ions, the optical depth of the environment becomes important. If the environment is optically thin \citep[Case A;][]{Osterbrock1989}, the ionizing photons were allowed to escape, while in the optically thick case (Case B), the photons were reabsorbed by a nearby neutral atom, which has the effect of lowering overall recombination rate. 
We included these effects for hydrogen, which has the highest number density and provides most of the free electrons.
To best estimate the appropriate case for each run, at each time step we calculated the optical depth as,
\be
\tau_H = \bar{X}_{H} \bar{\rho} \sigma_{\nu} L_{\rm box} / m_H,
\ee
where $\bar{X}_H$ is the global neutral hydrogen mass fraction, $\bar{\rho}$ is the mean ambient density, $\sigma_{\nu} = 3.3\times10^{-18}$ cm$^{2}$ is the photoionization cross section for hydrogen, and m$_H$ is the mass of hydrogen. The recombination rate for hydrogen is then
\be
k_{\rm rec} = e^{-\tau_H} k_A + (1.0 - e^{-\tau_H}) k_B,
\ee
where k$_{rec}$ is the new recombination rate and k$_A$ and k$_B$ are the Case A and B recombination rates respectively, which differ by a factor $\lesssim 2$. When the optical depth is low, e$^{-\tau_H} \approx 1.0$ and we defaulted to the Case A rate. Conversely when the optical depth is large, e$^{-\tau_H} \approx 0.0$ and we used the Case B rate.   

 As noted above, we have defined the cooling rates for each species between 5000 and 10$^{8}$ K. In regions where the temperature is below this range, we turned off cooling while allowing the chemistry to evolve,  and we enforced an absolute temperature floor at 100 K. 

\begin{table*}
\small
\caption{Summary of the models. Those models with the appended \_C denote runs made with thermal conduction while those appended with \_High denote high resolution models with twice the normal resolution.  Those denoted with an asterisk denote models that result in thermal runaway.}
\begin{threeparttable}
%\centerline{
%\scalebox{0.9}{
\begin{tabular}{lllllllllllll}
\hline
Name	  & $\bar \rho$ &	Column & $\sigma_{1D}$ & T$_{\rm MW}$ &  T$_{\rm VW}$ & M$_{\rm MW}$ & M$_{\rm VW}$ & $\bar{s}$ & $\sigma_s$ & $s_{\rm skew}$ & $s_{\rm kurt}$ & $\sigma_{\rm s,exp}$\\ 
\hline
\hline
N1E14\_S6*      & 7e-31 & 1.1e+14 & 5.8   & 5.52  & 5.51  & 0.29  & 0.30  &  0.004 & 0.05  & -0.60 & 1.01  & 0.15  \\ 
N1E15\_S6       & 7e-30 & 1.1e+15 & 5.8   & 2.55  & 2.54  & 0.29  & 0.30  &  0.005 & 0.05  & -0.50 & 1.74  & 0.15  \\ 
N1E16\_S6 [Low] & 7e-29 & 1.1e+16 & 5.8   & 1.49  & 1.48  & 0.51  & 0.51  &  0.005 & 0.13  & -0.84 & 1.90  & 0.25  \\ 
N1E17\_S6       & 7e-28 & 1.1e+17 & 5.8   & 1.26  & 1.24  & 0.61  & 0.62  &  0.001 & 0.16  & -0.79 & 1.48  & 0.30  \\ 
N1E16\_S12*     & 7e-29 & 1.1e+16 & 11.5  & 26.85 & 26.85 & 0.21  & 0.21  &  0.002 & 0.02  & -0.54 & 1.95  & 0.10  \\ 
N3E16\_S12      & 2e-28 & 3.0e+16 & 11.5  & 7.01  & 6.99  & 0.44  & 0.45  &  0.006 & 0.08  & -0.60 & 1.43  & 0.22  \\ 
N1E17\_S12      & 7e-28 & 1.1e+17 & 11.5  & 1.30  & 1.21  & 1.06  & 1.15  & -0.061 & 0.46  & -0.57 & 0.62  & 0.53  \\ 
N3E17\_S12      & 2e-27 & 3.0e+17 & 11.5  & 1.14  & 1.02  & 1.26  & 1.39  & -0.119 & 0.61  & -0.83 & 2.16  & 0.63  \\ 
N7E16\_S20*     & 5e-28 & 7.6e+16 & 20.2  & 34.05 & 33.99 & 0.36  & 0.37  &  0.006 & 0.06  & -0.60 & 1.95  & 0.18  \\ 
N1E17\_S20 [Med]& 1e-27 & 1.5e+17 & 20.2  & 5.54  & 5.31  & 0.91  & 0.96  & -0.021 & 0.31  & -0.68 & 1.16  & 0.46  \\ 
N7E17\_S20      & 5e-27 & 7.6e+17 & 20.2  & 1.01  & 0.82  & 2.39  & 2.95  & -0.381 & 0.98  & -0.19 & -0.27 & 1.07  \\ 
N1E18\_S20      & 1e-26 & 1.5e+18 & 20.2  & 0.87  & 0.69  & 2.78  & 3.56  & -0.497 & 1.11  & -0.21 & -0.11 & 1.20  \\ 
N4E17\_S35*     & 3e-27 & 4.6e+17 & 34.6  & 176.12 & 175.95 & 0.31  & 0.31  & 0.004 & 0.05  & -0.49 & 1.34  & 0.16  \\ 
N1E18\_S35      & 1e-26 & 1.5e+18 & 34.6  & 1.12  & 0.89  & 4.13  & 5.25  & -0.798 & 1.42  & -0.33 & 0.11  & 1.44  \\ 
N4E18\_S35      & 3e-26 & 4.6e+18 & 34.6  & 0.90  & 0.73  & 4.63  & 6.49  & -0.877 & 1.52  & -0.37 & 0.16  & 1.56  \\ 
N1E19\_S35      & 1e-25 & 1.5e+19 & 34.6  & 0.83  & 0.64  & 5.32  & 7.65  & -0.844 & 1.43  & -0.17 & -0.10 & 1.66  \\ 
N3E18\_S58*     & 2e-26 & 3.0e+18 & 57.7  & 67.55 & 64.53 & 0.91  & 0.96  & -0.005 & 0.32  & -0.54 & 0.73  & 0.46  \\ 
N1E19\_S58 [High] & 7e-26 & 1.1e+19 & 57.7  & 1.06  & 1.09  & 7.98  & 9.83  & -1.196 & 1.77  & -0.32 & 0.07  & 1.80  \\ 
N3E19\_S58      & 2e-25 & 3.0e+19 & 57.7  & 0.86  & 0.71  & 8.79  & 13.55 & -1.375 & 1.93  & -0.32 & -0.16 & 1.96  \\ 
N1E20\_S58      & 7e-25 & 1.1e+20 & 57.7  & 0.77  & 0.64  & 10.64 & 14.51 & -1.441 & 2.01  & -0.46 & 0.33  & 2.00  \\ 
N1E16\_S6\_High   & 7e-29 & 1.1e+16 & 5.8   & 1.49  & 1.48  & 0.45  & 0.45  & 0.009 & 0.10  & -0.44 & 1.12  & 0.22  \\ 
N1E16\_S6\_C    & 7e-29 & 1.1e+16 & 5.8   & 1.50  & 1.49  & 0.49  & 0.49  & 0.007 & 0.12  & -1.11 & 3.35  & 0.24  \\ 
N1E17\_S20\_High  & 1e-27 & 1.5e+17 & 20.2  & 5.66  & 5.43  & 0.89  & 0.94  & -0.021 & 0.32  & -0.68 & 0.90  & 0.45  \\ 
N1E17\_S20\_C   & 1e-27 & 1.5e+17 & 20.2  & 5.61  & 5.35  & 0.88  & 0.93  & -0.028 & 0.34  & -0.64 & 0.84  & 0.44  \\ 
N1E19\_S58\_High  & 7e-26 & 1.1e+19 & 57.7  & 0.95  & 0.81  & 7.68  & 10.64 & -1.173 & 1.74  & -0.23 & -0.22 & 1.84  \\
N1E19\_S58\_C   & 7e-26 & 1.1e+19 & 57.7  & 1.07  & 1.14  & 7.36  & 10.41 & -1.303 & 1.94  & -0.57 & 0.66  & 1.83  \\ 
\end{tabular}
%}}
\begin{tablenotes}[flushleft]
\item  \textbf{Notes.} $\bar \rho$ is the mean density in units of gm cm$^{-3}.$ Column is the column density in units of cm$^{-2}.$ $\sigma_{1D}$ is the 1-D velocity dispersion in units of km/s. $T_{\rm MW}$ and $T_{\rm VW}$ are the mass-weighted and volume-weighted temperatures in units of $10^4$ K. $M_{\rm MW}$ and $M_{\rm VW}$ are the mass-weighted and volume-weighted turbulent Mach numbers. $\bar s$ is the volumed averaged value of $s \equiv \ln \rho /\bar \rho.$  $\sigma_s,$ $s_{\rm skew}$  and $s_{\rm kurt}$ are the rms, skewness and kurtosis excess of the volume-weighted probably density function of $s$.  $\sigma_{\rm s, exp}$ is the expected variance from Eqn.~\ref{sigmam}.
\end{tablenotes}
\end{threeparttable}
\label{tab:simruns}
\end{table*}

\begin{figure*}
\begin{center}
\includegraphics[trim=0.0mm 0.0mm 0.0mm 0.0mm, clip, scale=0.85]{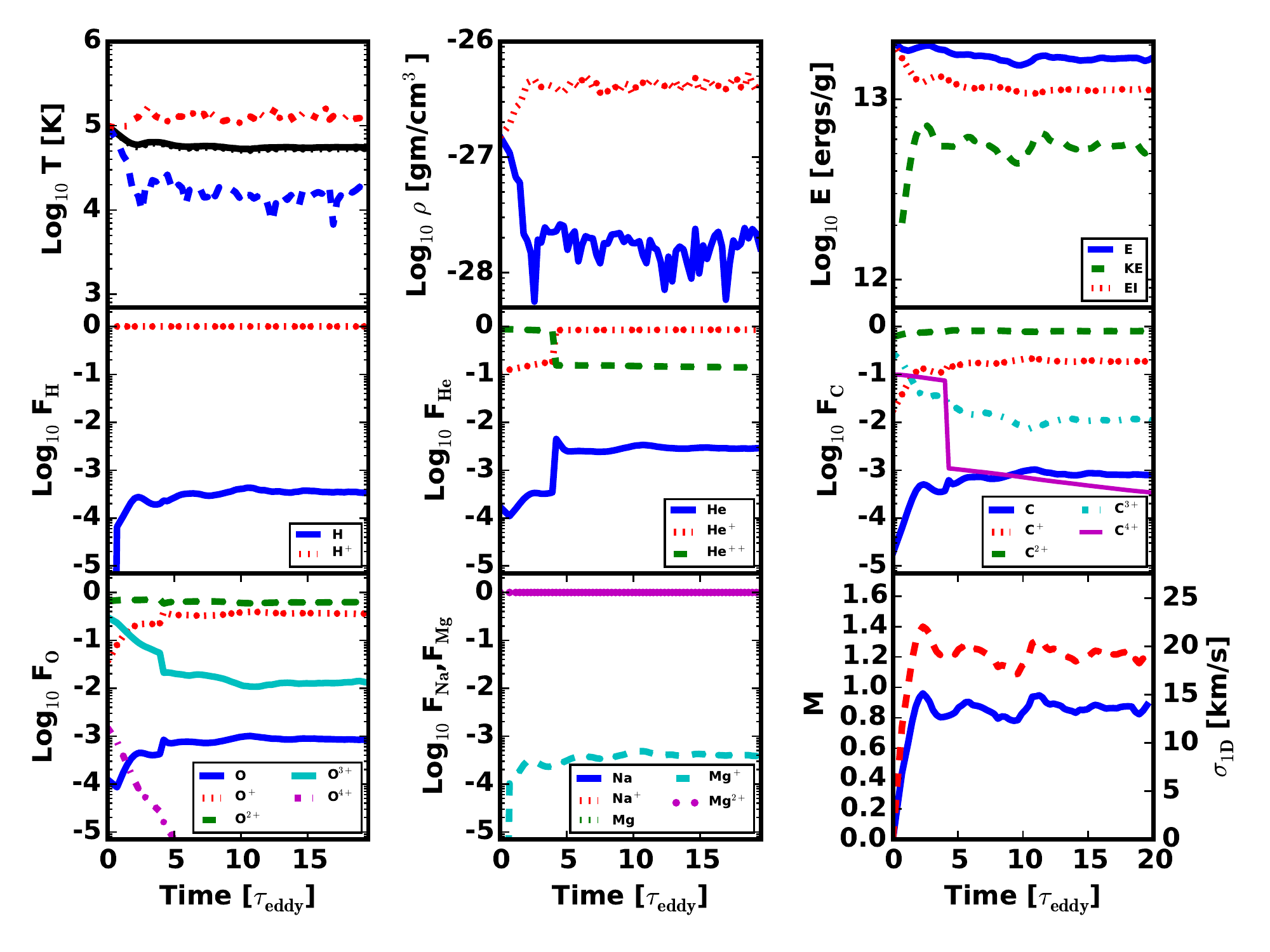}
\caption{Hydrodynamical and chemical evolution of N1E17\_S20. {\it Top Left:} The (red) dash-dotted line shows the maximum temperature, the (blue) dashed line shows the minimum temperature, the (black) dotted line shows the volume weighted temperature, and the (black) solid  shows the density weighted temperature. {\it Top Center:} The (blue) solid line shows the minimum density and the (red) dotted line shows the maximum density. {\it Top Right:} The (blue) solid line shows the total energy, the (red) dotted line shows the internal energy, and the (dashed) green line shows the kinetic energy.  {\it Middle Row:} The fractional abundances of the chemical species where each panel shows a different element as noted by the panel legends. {\it Bottom Left:} Same as middle row for Oxygen. {\it Bottom Center:} Same as the middle row. For convenience we have combined Na and Mg. {\it Bottom Right:} The (blue) solid, left axis line shows the Mach number of the simulation and the (red) dashed, right axis line is the 1D velocity dispersion.} 
\label{evo}
\end{center}
\end{figure*}

\begin{figure*}
\begin{center}
\includegraphics[trim=0.0mm 0.0mm 0.0mm 0.0mm, clip, scale=0.85]{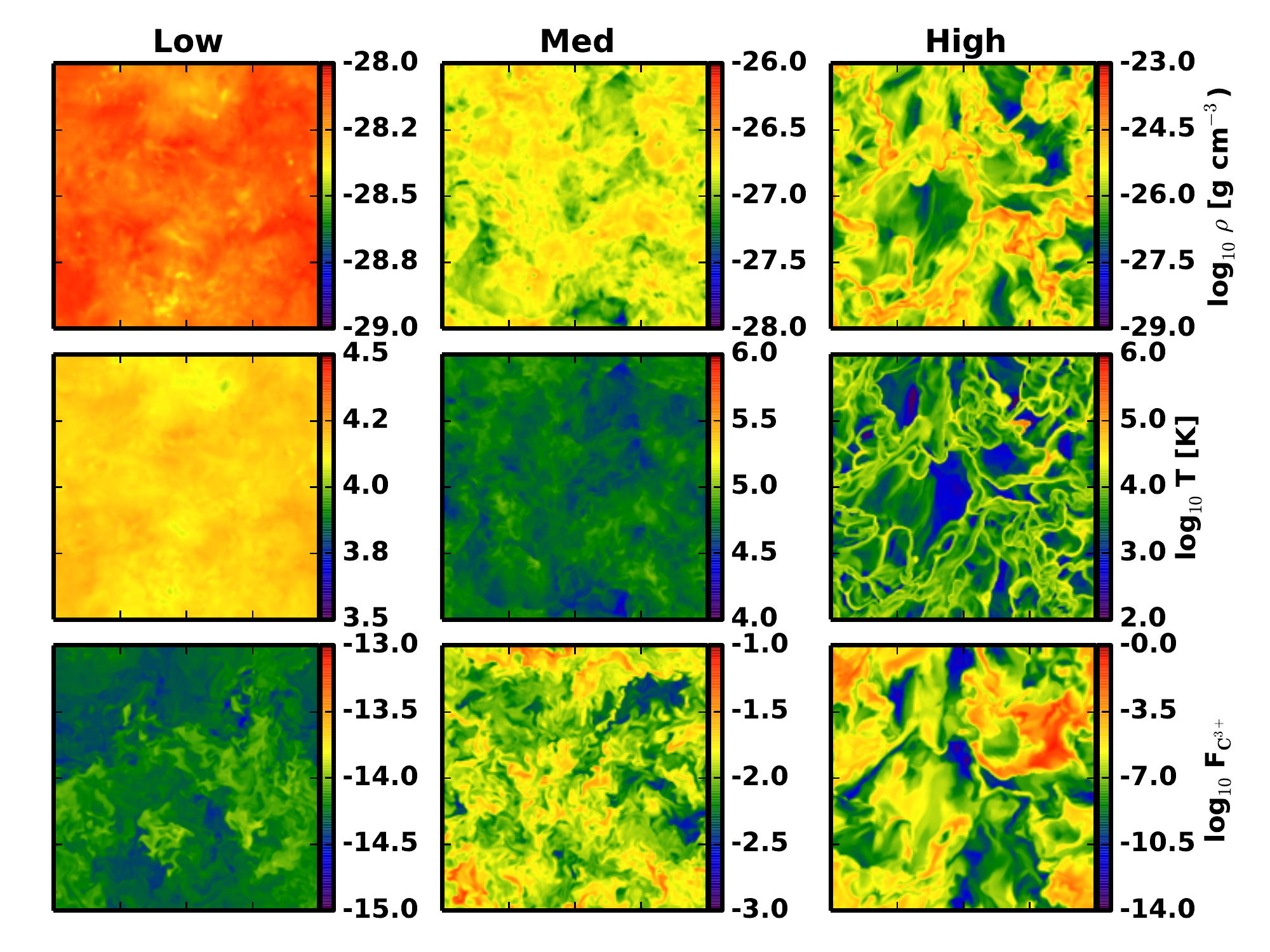}
\caption{ Density (top row), temperature (middle row) and C$^{3+}$ fraction (bottom row) on a fixed slice taken from our three representative models, Low (left column), Med (center), and High (right)}
\label{2dpanel}
\end{center}
\end{figure*}

\section{Results}

\subsection{Model Parameters}

Our goal is to  study the steady-state ionization structure of a gas that is being stirred with 1D velocity dispersions between 6 and 58 km $\rm s^{-1}$ (corresponding to 3D velocity dispersions between 10 and 100 km $\rm s^{-1}$), over a wide range of densities. In particular, we are interested in  cases in which the heating from the turbulent stirring is balanced by atomic cooling.

The parameter space spanned by our simulations is greatly simplified by the dependencies of turbulent decay and cooling on density and length scale. For a given steady state, the average turbulent energy dissipation rate per unit volume, or consequently, the average heating rate per unit volume is
\be
\bar \Gamma = \bar \rho \sigma^3 /L \ {\rm erg\ s^{-1}\ cm^{-3} },
\ee
where $\sigma$ is the velocity dispersion, $\bar \rho$ is the average density, and $L$ is the driving scale of the turbulence. Conversely, the average cooling rate per unit volume is
\be
\bar \Lambda \propto \bar \rho^2 \bar \lambda(T)/ (\mu m_{H})^2 \ { \rm erg\ s^{-1}\ cm^{-3}},
\ee
where $\mu$ is the average particle mass and $\bar \lambda(T)$ is the average cooling rate. Equating these two terms gives
\be
\bar \lambda(T) \propto \sigma^3 /(L \rho),
\ee
or in other words that the overall thermal distribution of the gas is only dependent on $\sigma$ and the column density. For the velocity dispersions we are interested in, for which cooling balances heating, this column density ranges between 10$^{16}$ and 10$^{20}$ cm$^{-2}$. 

Furthermore, because collisional ionization and recombination rates are also proportional to density squared, the ratio of the eddy turnover time defined here as $t_{\rm eddy}  = L_{\rm box}/2 \sigma,$ to the density-temperature averaged recombination time, $t_{\rm rec},$ is also purely a function of $\sigma$ and column density, with $t_{\rm rec} \geq t_{\rm eddy}$ for many species.  This means that many of the species of interest experience a wide range of conditions within a recombination time, which, as we shall see below in more detail,  can lead to abundance ratios that can not be predicted from local collisional equilibrium.  

A summary of the simulation runs is given in Table~\ref{tab:simruns}, with the name of each run referring to its column density and 1D velocity dispersion.
In the analysis that follows, we have chosen three representative models to describe in detail: N1E16\_S6, N1E17\_S20, and N1E19\_S58, which span a range of densities and velocity dispersions,  We will refer to these as Low, Medium, and High respectively, as they represent cases with progressively higher velocities and  turbulent Mach numbers.
Figure~\ref{evo} shows the hydrodynamic and chemical evolution of N1E17\_S20 in units of the eddy turnover timescale. The effect of the chemical kick is apparent at $t_{\rm eddy} \approx 5$ as the large jump in the fractional abundances of each species, e.g.,  $F_{C} = F_{C_i} / \sum_i F_{C_i}$.  After this adjustment, the species quickly find a global steady state that is distinct from instantaneous collisional ionization equilibrium. To ensure that the chemical kick did not introduce any bias in our model, we ran model N1E17\_S20 without the kick. The statistical and atomic properties compared very well between these two models, although the run without the kick took over four times longer to reach a final steady-state. 

Figure \ref{2dpanel} shows the density, temperature, and the fraction of carbon in the commonly observed C$^{3+}$ state on slices extracted from each of our representative models after they have reached a global steady state. Here we see that while density and temperature variations are small at low  turbulent Mach numbers, the more highly turbulent simulations display a wide range of temperatures and densities, which are only weakly correlated with each other.   Similarly, the distribution of the fraction of C$^{3+}$, which has a recombination time that is comparable to $t_{\rm eddy}$, displays features that  sometimes follow the temperature distribution, but sometimes show substantial variations, indicating that F$^{\rm C^{3+}}$ is often very different than would be estimated from local collisional ionization equilibrium. Below we study each of these effects in turn.

\begin{figure*}
\begin{center}
\includegraphics[trim=0.0mm 0.0mm 0.0mm 0.0mm, clip, scale=0.8]{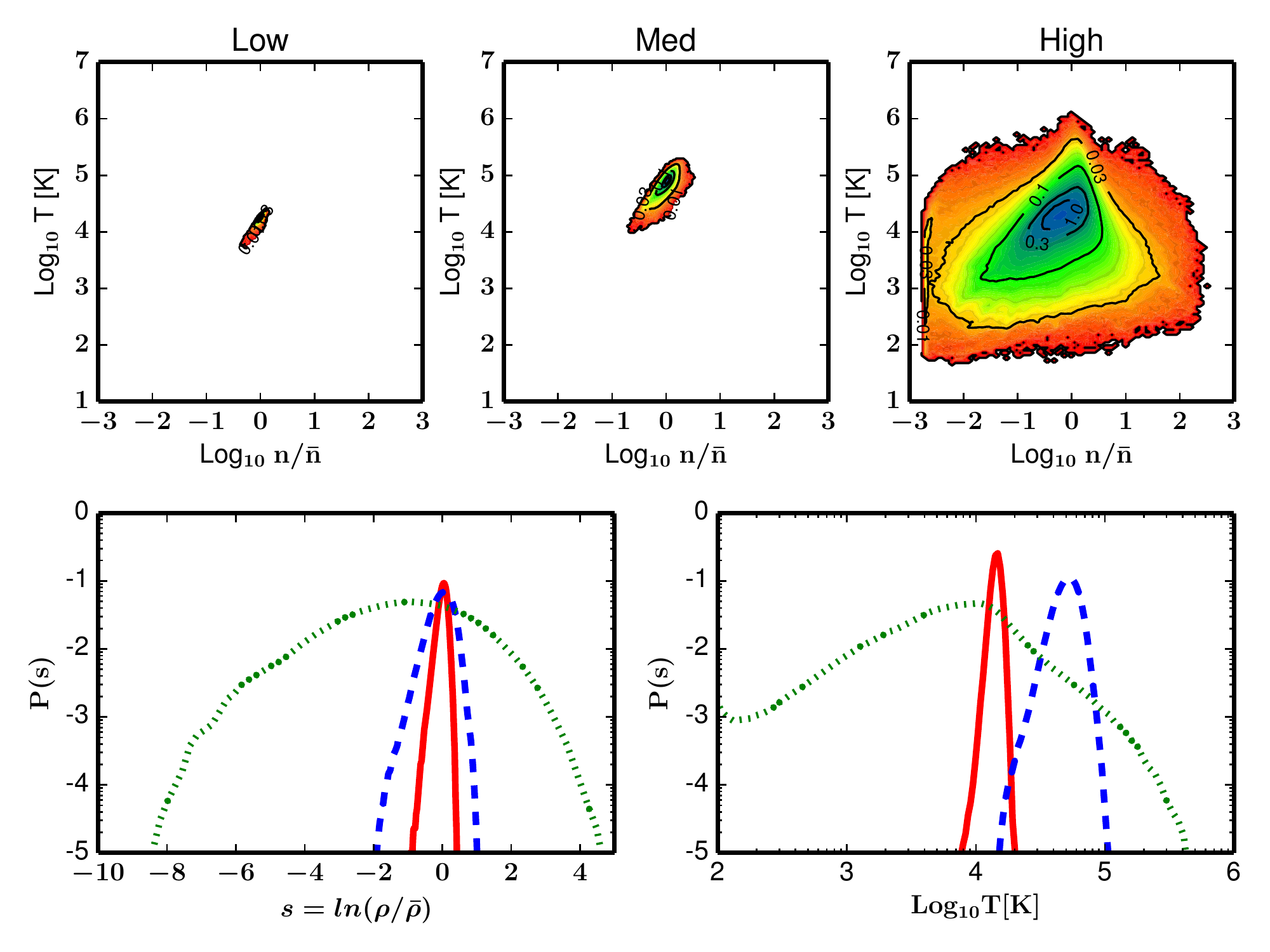}
\caption{ {\it Top:} Two-dimensional probability density function for simulations Low (Left), Med (Center), and High (Right).  Temperature is given along the $y$-axis in units of K and the number density is given along the $x$-axis. Here we have normalized by the average number density, $\bar{n}$, which corresponds to -4.43, -3.02, and -1.94 for Low, Med, and High respectively. All contours are labeled by their values relative to the PDF bin with the most mass. 
{\it Bottom Left:} One-dimensional probability density functions for run Low, Med, and High. The $x$-axis gives the logarithmic density. Low is shown by the (red) solid line, Med is shown by the (blue) dashed line, and High is shown by the (green) dotted line. The density variance increases as the stirring increases. Also note that the profiles are not symmetric with longer tails toward lower densities. {\it Bottom Right:} One-dimensional temperature PDFs for Low, Med, and High. The $x$-axis gives the temperature. The lines have the same meanings as in the other panel.}
\label{spdf}
\end{center}
\end{figure*}

\begin{figure}
\begin{center}
\includegraphics[trim=5.5mm 0.0mm 0.0mm 0.0mm, clip, scale=0.45]{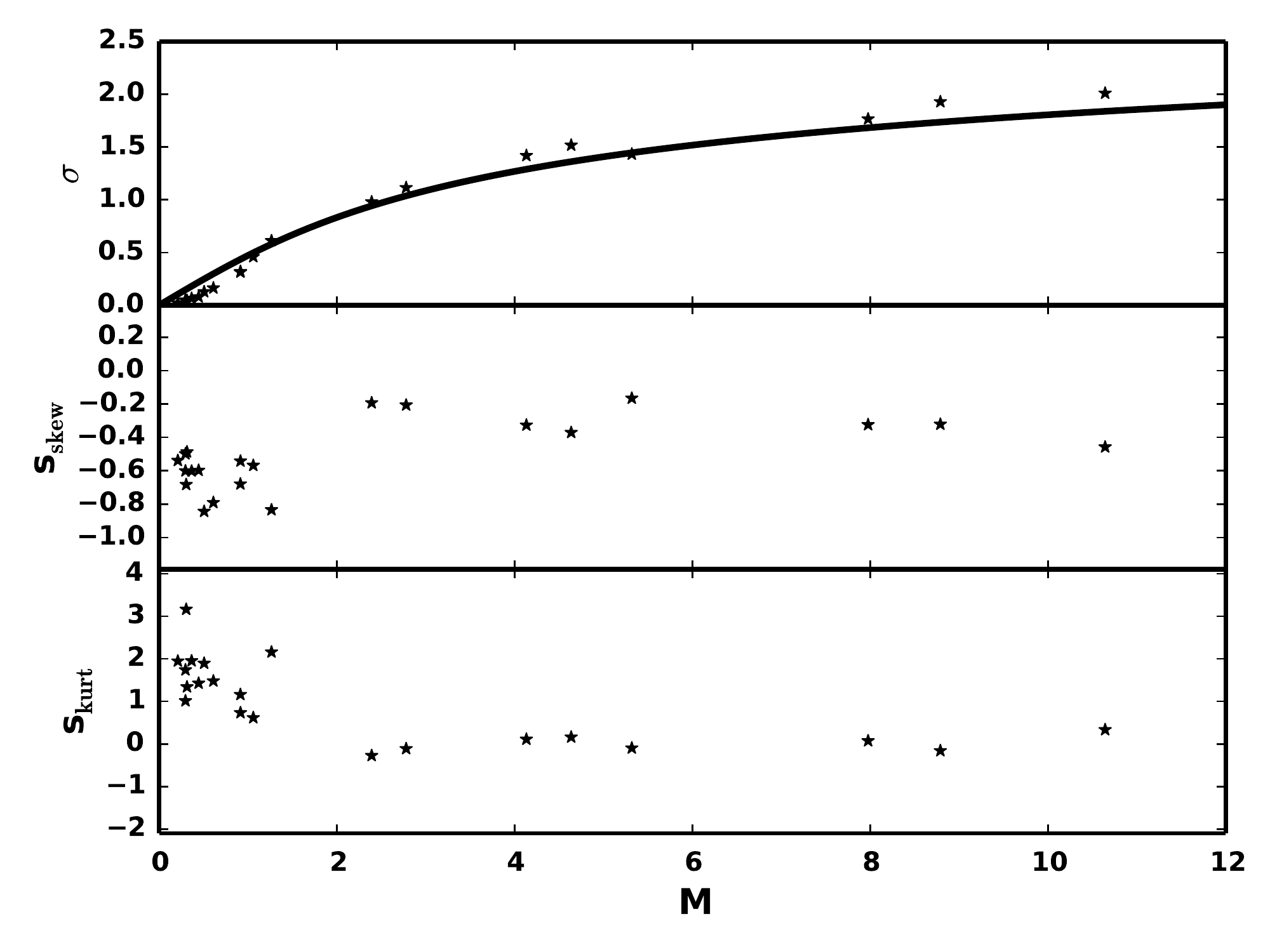}
\caption{Statistical measures of the $s \equiv \ln(\rho/\bar \rho)$ PDF for each model. The $x$-axis is the density weighted Mach number. {\it Top}: The points show the variance in $s$ for each model, and the solid line shows the expected relation from eqn~\ref{sigmam} with $b=0.5.$ Note that for supersonic flows the solid line matches very well with the points, but for subsonic flows the line does not match the data well. {\it Middle}: The skewness for each model. Note that for all models, the skewness is negative, which indicates longer tails toward lower densities. {\it Bottom}: The kurtosis. At subsonic Mach numbers most models show a more peaked distribution while at supersonic Mach numbers the distribution is either nearly gaussian or slightly flatter than gaussian.}
\label{sigma}
\end{center}
\end{figure}

\subsection{Probability Density Functions}

The top panel of Figure \ref{spdf} shows the two-dimensional volume-weighted probability density functions, which quantify the fraction of the volume in the simulations located at various temperatures and densities. Consistent with the expectations from Figure~\ref{2dpanel}, the gas in the Low simulation has a very small spread of temperatures and densities. In the Medium simulation, on the other hand, the temperatures and densities span roughly an order of magnitude. In this case, for which the (mass weighted) turbulent Mach number is  $M_{\rm MW} = 0.91,$ $T$ is roughly proportional to $n^{-1},$ as would occur exactly in a constant pressure medium, but  occurs here with a considerable spread, consistent with the presence of transonic motions.   Finally, in the High case, with a Mach number of $M_{\rm MW} = 8.0,$ the densities and temperatures span nearly six orders of magnitude and four orders of magnitude, respectively. Furthermore, these quantities are largely uncorrelated with each other, as would be expected in a medium filled with strong, supersonic shocks.

While our simulations are the first to include full rate equations and cooling for a large number of atomic species, several studies of isothermal, supersonic turbulence have shown that the gas approximates a lognormal distribution \citep{Vazquez1994,Padoan1997,Klessen2000,Ostriker2001,Yi2003,Kritsuk2007,Federrath2008,Lemaster2008,Schmidt2009,Federrath2010,Glover2010,Padoan2011,Collins2011,Price2011,Molina2012}, defined as
\be
p_s\ ds = \frac{1}{\sqrt{2 \pi \sigma_s^2}} {\rm exp} \left[-\frac{(s-s_0)^2}{2\sigma_s^2}\right],
\ee
where $s$ is the logarithmic density, $s \equiv {\rm ln}(\rho/\bar \rho).$

The mean logarithmic density in this case is related to the standard deviation as $s_0 = -\sigma_s^2/2$. The density variation at a particular location is produced by the successive passage of shocks with mach numbers independent of the local density, which gives a physical explanation for the lognormal density distribution. For an isothermal distribution, the variance of the logarithmic density corresponds to
\be
\sigma_{s}^2 = {\rm ln}( 1 + b^2 M^2 ),
\label{sigmam}
\ee
where $b$ is a constant that depends on the forcing that drives the turbulence. \cite{Federrath2008} showed that $b=1$ for purely compressive, $\nabla \times {\bf F} = 0,$ forcing, while  $b=1/3$ for purely solenoidal, $\nabla \cdot {\bf F} = 0$, forcing.

The bottom, left panel of Figure~\ref{spdf} shows the probability density functions of the logarithmic density, which take on a gaussian profile for runs Low, Med, and High. As expected, the width of the profile increases as the steady state Mach number increases, with Low having the smallest width and High having the largest. In Table~\ref{tab:simruns} we show several important statistical quantities in terms of the logarithmic density for each run. In particular, we calculate the first four moments of the logarithmic density distribution: the mean, variance, skewness, and kurtosis. 

The mean and variance have the usual definitions, while the skewness, $\left< (s-s_0)^3 \right>/ \sigma^{3/2}$, measures the symmetry of the distribution.  A positive skewness denotes a distribution that has a long tail toward higher densities while a negative skewness represents a long tail toward lower densities. The kurtosis, $\left< (s-s_0)^4 \right>/ \sigma^{2}$ quantifies the peakedness of the distribution and measures the importance of the tails versus the peak. A larger kurtosis represents a flatter distribution while a lower value denotes a narrower distribution, with a gaussian distribution having a kurtosis equal to 3. In Table~\ref{tab:simruns} we give the kurtosis excess where this factor of 3 is subtracted. 

Figure~\ref{sigma} shows the variance, skewness, and kurtosis excess as a function of Mach number for each model. The solid line in the top panel shows the expected $\sigma_s$-M relation from equation~\ref{sigmam}, with $b=0.5$. We find that for supersonic flows the models match very well with equation~\ref{sigmam} but find a significant mismatch for subsonic flows. A least-squares fit of equation~\ref{sigmam} to our points suggests a good fit to the $M \gtrsim 1$ points with $b=0.53$. However, this does not match the expectation from \cite{Federrath2008} who suggest a value of $b=1/3$ for purely solenoidal driving, for isothermal turbulent models. This difference may simply be due to the non-isothermal nature of our models. The local variation in temperature allows for a larger spread in density than in isothermal models. 

The right panel of Figure~\ref{spdf} show the 1-D temperature PDFs for runs Low, Med, and High. As expected, the width of the PDF grows with the strength of the stirring, meaning that isothermality is not a good approximation for our high mach number runs,  which may help explain the discrepancy in $b.$  Note also the large temperature spread causes a few cells in run High to reach the temperature floor,  due to the extreme $PdV$ work done by the stirring, but this is only a minor effect.

The top row of Figure~\ref{spdf} shows the two dimensional PDFs of density and temperature for runs Low, Med, and High. For the Low case, the slight stirring has only produced very minor density contrasts, leaving a mostly uniform medium. However, there is a slight tail towards lower density and lower temperature. This leads to a highly concentrated phase diagram where essentially all the mass has the same temperature. This can also be seen in the left column of Figure~\ref{2dpanel} where we show slices of density, temperature, and the abundance of \cthree.  

A similar trend is seen for run Med in the center panel of Figure~\ref{spdf} and the center column of Figure~\ref{2dpanel}. Although, as mentioned below, the average mach number of the flow is higher than Low; turbulence has not produced any strong density contrasts. In fact, the density is almost always within a factor of 3 of the mean. As in the case for Low, the density and temperature slices are largely uniform.

Finally, the High case is quite different than either Low or Med. Large density and temperature contrasts are seen, due to the much higher final mach number ($M=8.0$). This produces a very wide density-temperature PDF as shown in the right panel of  Figure~\ref{spdf} with density and temperature spanning nearly six orders of magnitude and four orders of magnitude respectively. These contrasts are plainly seen in the right column of Figure~\ref{2dpanel}.

Finally, the bottom row of Figure~\ref{2dpanel} shows slices of F$_{\rm C^{3+}}$. In general the density and temperature distributions are very similar. For example, the density and temperature slices for High show that the complex density structure is largely mirrored in the temperature. However, this is not the case for \cthree.  In fact, there is only a weak relation between these quantities. As we discuss below, this highlights the need to follow the nonequilibrium atomic chemistry exactly when estimating atomic ionization states. 

\begin{figure*}
\begin{center}
\includegraphics[trim=0.0mm 0.0mm 0.0mm 0.0mm, clip, scale=0.72]{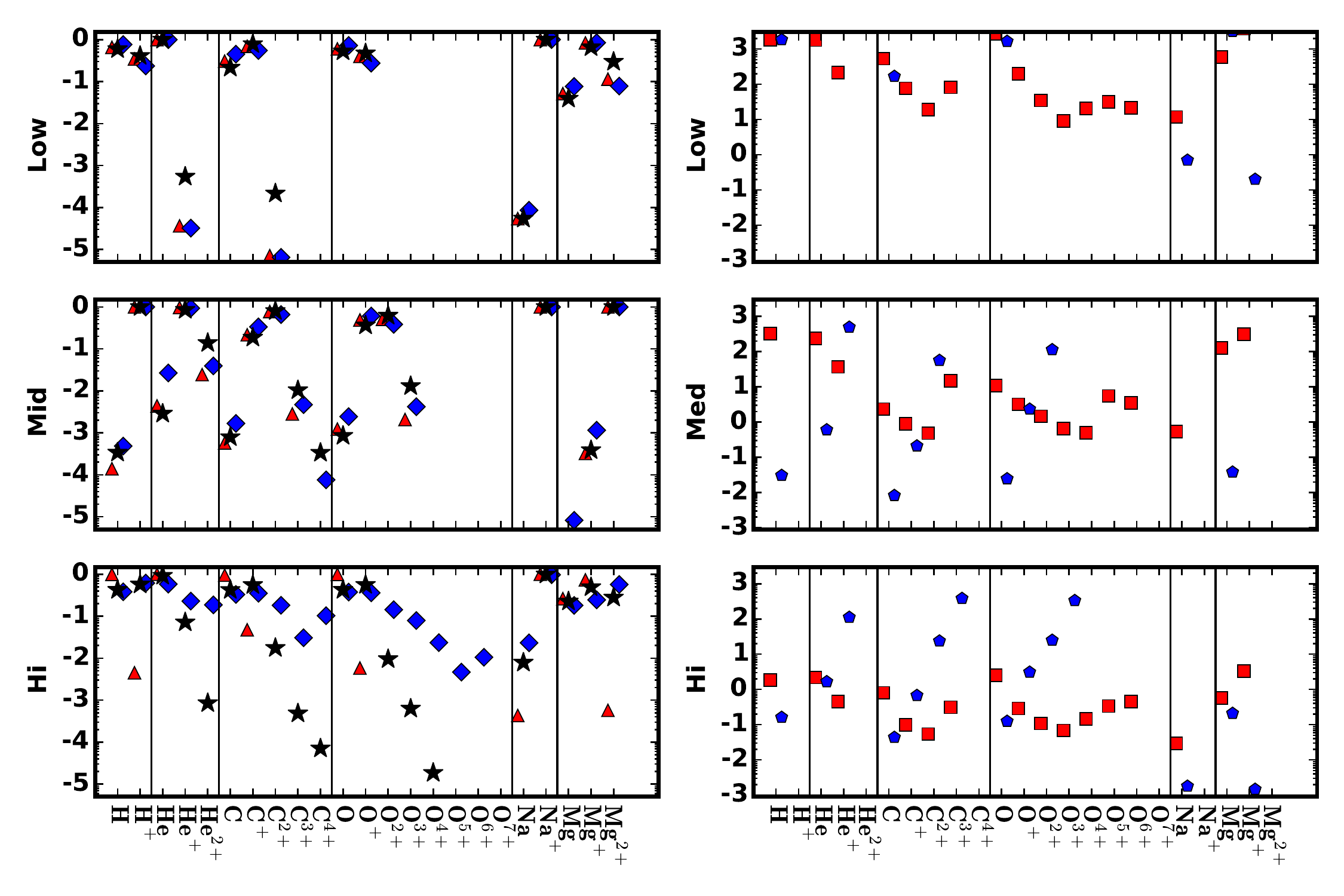}
\caption{ {\it Left:} Steady state species fractions for each of the representative runs. Each species is given along the $x$-axis, and the $y$ axis gives log10 of the fraction of the total mass in a given element that is found in a particular ionization state. The (red) triangles show $F_{\rm mean},$ the expected species fraction assuming collisional ionization equilibrium at the mass weighted mean temperature. The (blue) diamonds show  $F_{\rm TurbEq},$ the fraction assuming collisional ionization equilibrium using the temperature PDF. The (black) stars show, $F_{\rm Turb},$ the final global steady state fractions. {\it Right}: Comparison between the ionization and recombination timescales for each ion. The (red) squares and (blue) pentagons are the ratios of the recombination and ionization timescales with the eddy turnover time, respectively.}
\label{fracdiff}
\end{center}
\end{figure*}

\subsection{Ionization States}
\label{IS}

The large spread in both density and temperature in our simulations, particularly in the supersonic cases, naturally translates into a large spread in chemical species.  In fact, the instantaneous chemical makeup in a given cell is a strong function of not only the current temperature and density but its thermodynamic history. 

To get an idea of how the global steady state abundances depend on our full treatment of turbulence, cooling, and chemical reactions, we calculated approximate abundances using two approaches. For our first method, labeled $F_{\rm Mean}$, we assumed the full medium was in collisional ionization equilibrium at the mass weighted average temperature, and we calculated the ionization states for each species using Cloudy. This method quantifies the errors involved with assuming constant temperature and density conditions in a region with significant turbulent motions.  In our second estimate, labeled $F_{\rm TurbEq},$ we used the final temperature PDF to calculate a mass-weighted average abundance for each species, again using Cloudy to calculate collisional ionization abundances for the full range of conditions.   This method quantifies the errors involved with failing to track the nonequillibrium abundances accurately by including reaction rates within the simulations.

The left panel of Figure~\ref{fracdiff} shows the comparison between each of these estimates to the steady state abundances for Low, Med, and High. The name of each species is listed along the $x$-axis and the fractional species abundances are given along the $y$-axis.  These values are also listed in Table 3. As shown in above, for the Low simulation, with a mean temperature of $1.5 \times 10^4$K and a Mach number of $0.5$, the spread in the temperature-density PDF is small. Hence, the differences between the global species abundances and those from our two estimates are mostly minor, with most of the true values matching the simpler calculations within a factor of $\approx 2$ for most elements with significant abundances.  Two notable exceptions to this agreement are C$^{2+}$ for which the true abundance is $\approx 30$ times greater than the $F_{\rm Mean}$ and  $F_{\rm TurbEq},$ values, and He$^{+},$ which has a long recombination time, and a true abundance $\approx 20$ times greater than expected from more simple estimates.

\begin{table*}
\caption{Species fractions in our representative runs.}
\small
\centerline{
\scalebox{1.0}{
\begin{tabular}{lrrrrrrrrrr}
\hline
  & Low & Low & Low & Medium & Medium & Medium & High & High & High \\ 
  & $F_{\rm Mean}$  & $F_{\rm TurbEq}$ & $F_{\rm Turb}$ & $F_{\rm Mean}$  & $F_{\rm TurbEq}$ & $F_{\rm Turb}$
& $F_{\rm Mean}$  & $F_{\rm TurbEq}$ & $F_{\rm Turb}$ \\
\hline
\hline 
${\rm H}$       & -0.23  & -0.18  & -0.12  & -3.47  & -3.86  & -3.32  & -0.37  & -0.00  & -0.42   \\
${\rm H^+}$     & -0.39  & -0.46  & -0.63  & 0.00   & 0.00   & -0.00  & -0.24  & -2.35  & -0.21   \\
${\rm He}$      & -0.00  & 0.00   & 0.00   & -2.54  & -2.36  & -1.57  & -0.03  & 0.00   & -0.23   \\
${\rm He^{+}}$  & -3.26  & -4.44  & -4.49  & -0.07  & -0.01  & -0.03  & -1.14  & -8.03  & -0.64   \\
${\rm He^{2+}}$ & -17.28 & -30.00 & -30.00 & -0.86  & -1.61  & -1.40  & -3.07  & -30.00 & -0.73   \\
${\rm C}$       & -0.67  & -0.51  & -0.35  & -3.10  & -3.24  & -2.78  & -0.37  & -0.02  & -0.48   \\
${\rm C^{+}}$   & -0.11  & -0.16  & -0.26  & -0.73  & -0.66  & -0.47  & -0.25  & -1.32  & -0.45   \\
${\rm C^{2+}}$  & -3.67  & -5.15  & -5.19  & -0.10  & -0.11  & -0.18  & -1.75  & -9.79  & -0.74   \\
${\rm C^{3+}}$  & -14.19 & -30.00 & -18.41 & -1.98  & -2.55  & -2.33  & -3.31  & -30.00 & -1.51   \\
${\rm C^{4+}}$  & -18.56 & -30.00 & -30.00 & -3.47  & -5.94  & -4.12  & -4.15  & -30.00 & -0.99   \\
${\rm O}$       & -0.28  & -0.22  & -0.14  & -3.07  & -2.91  & -2.61  & -0.37  & -0.00  & -0.42   \\
${\rm O^{+}}$   & -0.33  & -0.40  & -0.56  & -0.44  & -0.31  & -0.22  & -0.25  & -2.24  & -0.44   \\
${\rm O^{2+}}$  & -7.62  & -8.81  & -8.49  & -0.21  & -0.29  & -0.41  & -2.02  & -15.66 & -0.84   \\
${\rm O^{3+}}$  & -17.97 & -30.00 & -30.00 & -1.88  & -2.68  & -2.38  & -3.20  & -30.00 & -1.10   \\
${\rm O^{4+}}$  & -18.92 & -30.00 & -30.00 & -5.92  & -30.00 & -6.05  & -4.73  & -30.00 & -1.63   \\
${\rm O^{5+}}$  & -18.72 & -30.00 & -30.00 & -11.67 & -30.00 & -11.69 & -6.21  & -30.00 & -2.33   \\
${\rm O^{6+}}$  & -18.85 & -30.00 & -30.00 & -13.14 & -30.00 & -30.00 & -7.53  & -30.00 & -1.98   \\
${\rm O^{7+}}$  & -19.68 & -30.00 & -30.00 & -20.89 & -30.00 & -30.00 & -15.17 & -30.00 & -10.37  \\
${\rm Na}$      & -4.26  & -4.27  & -4.07  & -5.86  & -6.29  & -6.11  & -2.10  & -3.37  & -1.63   \\
${\rm Na^{+}}$  & -0.00  & 0.00   & 0.00   & -0.00  & 0.00   & -0.00  & -0.00  & -0.00  & -0.01   \\
${\rm Mg}$      & -1.40  & -1.29  & -1.12  & -6.37  & -6.63  & -5.09  & -0.65  & -0.58  & -0.74   \\
${\rm Mg^{+}}$  & -0.18  & -0.08  & -0.07  & -3.41  & -3.49  & -2.94  & -0.31  & -0.13  & -0.61   \\
${\rm Mg^{2+}}$ & -0.53  & -0.94  & -1.11  & -0.00  & -0.00  & -0.00  & -0.55  & -3.24  & -0.24   \\          
\label{spectable}
\end{tabular}
}
}
\end{table*}

In the Med case, for which the mass-weighted mean temperature is $T_{\rm MW} = 5.5 \times 10^4$ and $M=0.92,$ the difference between the true values  and simple estimates becomes more pronounced.  Here the true abundances  of species such as He, C, C$^{+}$, C$^{3+}$, O$^{3+}$, and Mg, differ by a factor of $\approx 2$ from our simple estimates, while He$^{2+}$  and C$^{4+}$ are more abundant by $\approx 5$ than expected from either approximate approach.  Note also that there is no clear trend for the more complicated $F_{\rm TurbEq}$ estimate to be a better predictor than  the simple constant temperature approach.

Finally, in the High case, with $T_{\rm MW} = 1.1 \times 10^4$K and $M=8.0,$ the abundances of many species differ by several order of magnitude from simple expectations. For example, the abundance of H$^+$ is over two orders of magnitude higher than it is in the constant temperature models, but close to the $F_{\rm TurbEq}$ estimate. The abundance of He, on the other hand, is very near the value expected from  $F_{\rm Mean}$ and roughly twice the $F_{\rm TurbEq}$  estimate, while and He$^{2+}$ is many orders of magnitude higher than expected by the $F_{\rm Mean}$  estimate, but still over 200 times less abundant than expected in the  $F_{\rm TurbEq}$  estimate.  Similarly, C$^{2+}$, C$^{3+}$, C$^{4+},$ O$^{2+},$ O$^{3+}$, and O$^{4+}$ which are almost completely absent in the $F_{\rm Mean}$, are all present at significant levels, although less so than expected by $F_{\rm TurbEq}$. On the other hand, Mg$^+$ is present at a level much less than expected by  the $F_{\rm Mean}$  estimate, but more than predicted by $F_{\rm TurbEq}$.   These results not only show that  the presence of turbulence can introduce significant abundance differences when $M \approx 1$, and change the abundances completely when $M \gtrsim 1$, but that these differences can only be predicted by full nonequilibrium calculations such as the ones carried out here. 

To get an idea of the cause of the large discrepancy between carbon and oxygen in the High case, we compare the ionization and recombination timescales of these ions. This is shown in the right panel of Figure~\ref{fracdiff}. We find that while the recombination time is short compared to the eddy turnover, the ionization timescale for these ions is long. Physically, this suggests that even though a parcel of gas is heated due to the crossing shocks, the ions do not have sufficient time to fully ionize. This explains the lower abundances found for $F_{\rm TurbEq}$ compared to $F_{\rm Mean}$ for \ctwo through \cfour\ and O$^{3+}$ through \oseven\ for case High. 

\begin{figure*}
\begin{center}
\includegraphics[trim=0.0mm 0.0mm 0.0mm 0.0mm, clip, scale=0.75]{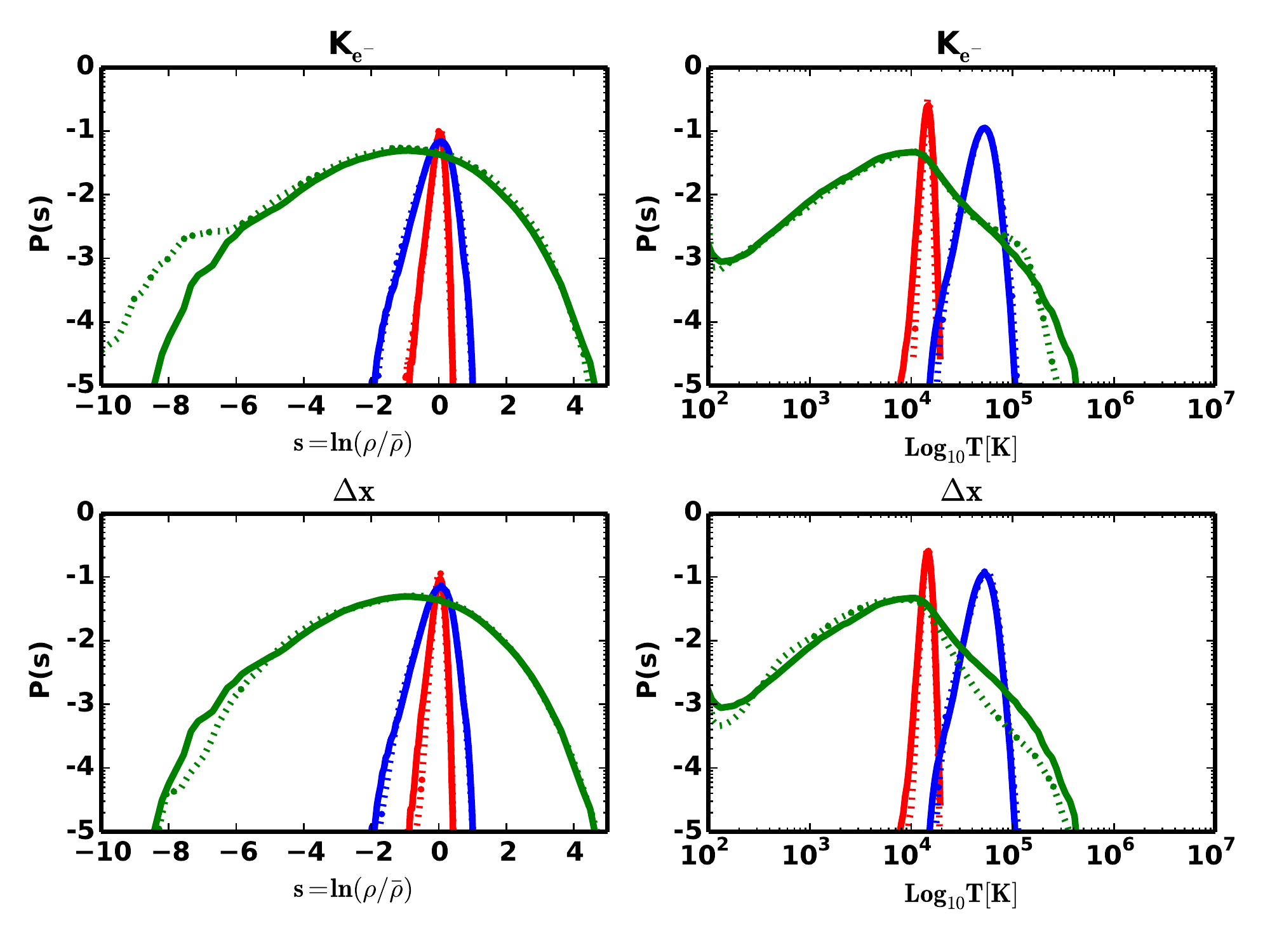}
\caption{ Effect of resolution and thermal conduction.  {\it Top Left}: Effect of electron conduction on the logarithmic density PDF. The solid lines are the fiducial models while the dashed lines show the model with conduction. {\it Top Right}: Effect of electron conduction on the temperature PDF.  {\it Bottom Left}: Comparison between our fiducial models and models with twice the base resolution on the logarithmic density PDF. {\it Bottom Right}: Comparison between our fiducial models and models with twice the base resolution on the temperature PDF. The solid lines are the fiducial models while the dashed lines show the higher resolution models. }
\label{compruns}
\end{center}
\end{figure*}

\subsection{Thermal Conduction and Resolution}

In our simulations described above, energy transport is purely by advection.  Yet even in a supersonic medium, the large difference in the thermal velocities of the electrons and the ions can lead to significant energy transport through conduction.  In the case in which many collisions take place over the scale length for temperature variations, the thermal conductivity is given by
\be
\kappa =  \nu \left(\frac{m_p}{m_e}\right)^{1/2}  = 5.2 \times 10^{-5} T_{4}^{5/2} n^{-1} \qquad {\rm cm^2 \, s^{-1}},
\ee
where $\nu$ is the plasma viscosity and  $T$ is the temperature in units of $10^4$ K \citep{Braginskii1958, Spitzer1962}.
%{\rm Re} = 8.5 \times 10^5 (v/{\rm km \, s^{-1}}) (L/{\rm parsec}) (n/{\rm cm}^{-3}) (T/10^4 K)^{2.5}$
 Comparing this to the velocity dispersion and length scale of the turbulence gives $\sigma_{\rm 1D} L/\kappa \propto \sigma_{\rm 1D} n L $. Like the ratios of timescales discussed in section 4.1, this is dependent only on the column density and turbulent velocity.  In fact, the Reynolds number  of the medium also only depends on these two quantities, although in practice the numerical viscosity in our simulations is much greater than the true physical value.

In the case in which the scale length for temperature variations is smaller than the collisional mean free path of the electrons, the conduction becomes saturated.  The mean free path is $\lambda_i =  1.3 \times 10^{18} {\rm cm}^{-2} T_7^2/n_{i,c}$, where $n_{i,c}$ is the ion density, the saturated flux is $F_{\rm sat} \alpha_{\rm ele}  n_e  \sqrt{(k T)^3/m_e},$  where $\alpha_{\rm ele}$ is the electron conductivity flux-limiter coefficient,  $n_e$ is the electron number density, and $m_e$ is the mass of the electron \citep{Cowie1977}.  From these scalings we can see that  $L/\lambda_i$ and $F_{\rm sat}/ n kT (\sigma)$ are purely functions of column density and the temperature structure of the medium. This means that neither unsaturated or saturated conduction introduces a new parameter that must be spanned by our simulations.   

To test the impact of electron conduction, we re-ran our representative simulation using of the implicit SpitzerHighZ conductivity module within FLASH, a Larsen flux limiter, and an electron conductivity flux-limiter coefficient of $\alpha_{ele}=0.2$. The upper panel of Figure~\ref{compruns} compares the logarithmic density PDFs from our nominal models, with these conductions runs, from these runs, which were roughly twice as expensive in terms of computer time. For run Low the PDF profile is essentially unchanged from the nominal model, in particular with regards to the mean and variance of $s$. Table~\ref{tab:simruns} presents the statistical measures between our nominal runs and those with electron conduction.  There is a slight increase in the skewness and kurtosis which produces a slightly longer distribution to lower densities and slightly more peaked distribution. Run Med shows no substantial change in the kurtosis, but conduction does produce a slightly less peaked distribution. Run High shows a similar trend as the other models, with a slight increase in both the skewness and kurtosis, but now with a noticeable difference in the extreme low-density tail of the log density PDF.

The top right panel of Figure~\ref{compruns} shows the effect of electron conduction on the logarithmic temperature PDF. Like the density PDFs, the temperature PDFs are largely identical for Low and Med. Only for the High simulation is there a small decrease in the PDF at the highest temperatures at which conduction is most efficient, which is accompanied by an increase in the low-density end of the density pdf.  Thus it appears that in the highest Mach number cases, conduction is able to operate quickly enough to move a noticeable amount of energy out of the hottest cells, moving some of this energy into low density regions which are able to expand a bit more due to the increase in pressure.  However, these differences are small and suggest that electron conduction does not play a major role.

\begin{table*}
\caption{Effect of resolution and electron conduction of the final elemental abundances. Each row gives a difference species while each column gives a different simulation. \_High denotes models with twice the base resolution and those with \_C denote models with electron conduction.}
\small
\centerline{
\tabcolsep=0.11cm
\scalebox{1.0}{
\begin{tabular}{l|rrrrrrrrrr}
\hline
Col.\    (cm$^{-2}$)        & 1.1E16         & 1.1E16\_High     & 1.1E16\_C      & 1.5E17         & 1.5E17\_High     & 1.5E17\_C      & 1.1E19         & 1.1E19\_High     & 1.1E19\_C      \\
$\sigma_{1D}$  (km s$^{-1}$)  &  5.8           &  5.8           &  5.8           &  20.2          &  20.2          &  20.2          &  57.7          &  57.7          &  57.7          \\
\hline
\hline 
${\rm H}$       & -0.23          & -0.22          & -0.15          & -3.47          & -3.50          & -3.47          & -0.37          & -0.27          & -0.29           \\
${\rm H^+}$     & -0.39          & -0.40          & -0.52          & 0.00           & 0.00           & 0.00           & -0.24          & -0.34          & -0.31           \\
${\rm He}$      & -0.00          & -0.00          & -0.00          & -2.54          & -2.59          & -2.56          & -0.03          & -0.02          & -0.05           \\
${\rm He^{+}}$  & -3.26          & -3.34          & -3.71          & -0.07          & -0.07          & -0.07          & -1.14          & -1.30          & -0.95           \\
${\rm He^{2+}}$ & -17.28         & -17.39         & -17.92         & -0.86          & -0.81          & -0.84          & -3.07          & -3.14          & -2.48           \\
${\rm C}$       & -0.67          & -0.69          & -0.60          & -3.10          & -3.14          & -3.11          & -0.37          & -0.34          & -0.36           \\
${\rm C^{+}}$   & -0.11          & -0.10          & -0.13          & -0.73          & -0.75          & -0.74          & -0.25          & -0.27          & -0.26           \\
${\rm C^{2+}}$  & -3.67          & -3.60          & -3.80          & -0.10          & -0.09          & -0.09          & -1.75          & -2.26          & -1.92           \\
${\rm C^{3+}}$  & -14.19         & -14.15         & -18.30         & -1.98          & -1.91          & -1.95          & -3.31          & -3.77          & -3.20           \\
${\rm C^{4+}}$  & -18.56         & -18.41         & -18.13         & -3.47          & -3.18          & -3.63          & -4.15          & -4.26          & -3.59           \\
${\rm O}$       & -0.28          & -0.26          & -0.17          & -3.07          & -3.10          & -3.07          & -0.37          & -0.27          & -0.30           \\
${\rm O^{+}}$   & -0.33          & -0.35          & -0.50          & -0.44          & -0.46          & -0.44          & -0.25          & -0.34          & -0.31           \\
${\rm O^{2+}}$  & -7.62          & -8.05          & -13.47         & -0.21          & -0.20          & -0.21          & -2.02          & -2.38          & -1.94           \\
${\rm O^{3+}}$  & -17.97         & -17.82         & -17.52         & -1.88          & -1.85          & -1.93          & -3.20          & -3.62          & -3.02           \\
${\rm O^{4+}}$  & -18.92         & -18.73         & -18.44         & -5.92          & -5.85          & -6.01          & -4.73          & -4.81          & -4.54           \\
${\rm O^{5+}}$  & -18.72         & -18.57         & -18.29         & -11.67         & -11.25         & -11.84         & -6.21          & -5.96          & -6.09           \\
${\rm O^{6+}}$  & -18.85         & -18.66         & -18.49         & -13.14         & -10.67         & -16.48         & -7.53          & -7.05          & -6.88           \\
${\rm O^{7+}}$  & -19.68         & -19.17         & -19.82         & -20.89         & -19.84         & -20.17         & -15.17         & -17.10         & -17.03          \\
${\rm Na}$      & -4.26          & -4.29          & -4.29          & -5.86          & -5.85          & -5.85          & -2.10          & -2.12          & -2.22           \\
${\rm Na^{+}}$  & -0.00          & -0.00          & -0.00          & -0.00          & -0.00          & -0.00          & -0.00          & -0.00          & -0.00           \\
${\rm Mg}$      & -1.40          & -1.44          & -1.36          & -6.37          & -6.39          & -6.36          & -0.65          & -0.60          & -0.65           \\
${\rm Mg^{+}}$  & -0.18          & -0.17          & -0.12          & -3.41          & -3.44          & -3.41          & -0.31          & -0.24          & -0.24           \\
${\rm Mg^{2+}}$ & -0.53          & -0.54          & -0.72          & -0.00          & -0.00          & -0.00          & -0.55          & -0.76          & -0.69     \\      
\label{fdata0}
\end{tabular}
}
}
\vspace{0.3in}
\end{table*}

In order to simulate a large number of turbulent conditions for many eddy turnover times in the presence of chemistry at reasonable CPU cost, our simulations have adopted a relatively low resolution of $128^3.$   In this case, two-point statistics such as the velocity and density power-spectra and structure functions are not expected to be well reproduced  \citep[\eg][]{Falkovich1994, Kritsuk2007, Pan2010}, and we do not attempt to analyze them here.   On the other hand, the probability distribution function is a single-point quantity that is much easier to simulate at moderate resolution.   

As a check of convergence, the bottom left row of Figure~\ref{compruns} compares the PDF obtained in our representative simulations with a second set of simulations with the same set of physical parameters, but a resolution of $256^3.$  Again, we find that the resulting PDFs are nearly identical to the nominal runs and only in the High case is there a slight decrease in the high temperature tail and a slight dip in the low density tail. However, in terms of the mean and variance, the effect is minor, which suggests that our nominal model with $N$ = 128 is sufficient to produce the key one-point statistical quantities and obtain estimates of the species abundances.   

\begin{figure}
\begin{center}
\includegraphics[trim=0.0mm 0.0mm 0.0mm 0.0mm, clip, scale=0.56]{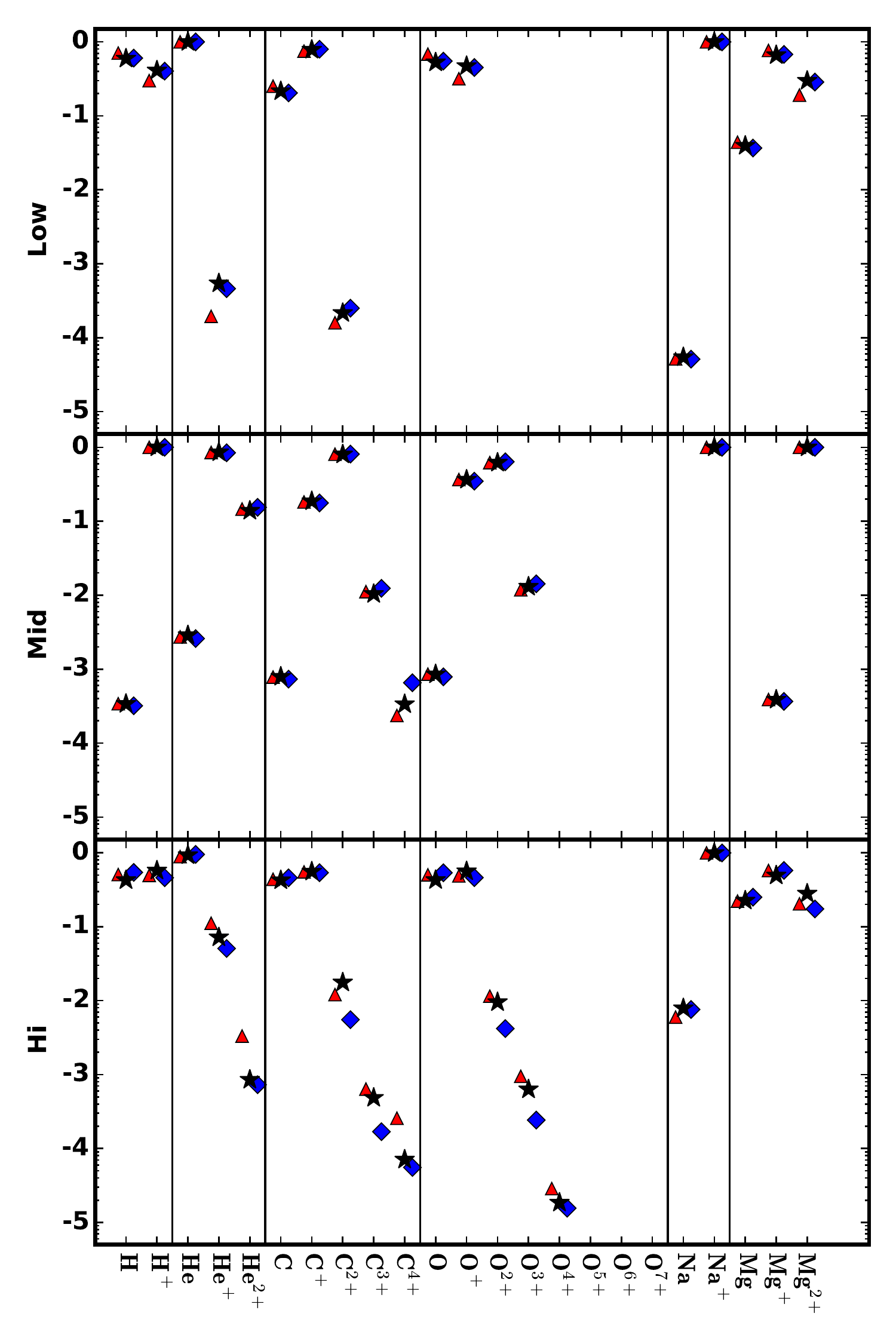}
\caption{Effect of electron conduction and resolution. Each species is given along the  $x$-axis. The (red) triangles show the steady state species fractions with electron conduction. The (blue) diamonds show the steady state species fractions with twice the base resolution. The (black) stars show the final global steady state.}
\label{fracphys}
\end{center}
\end{figure}

Finally, in Figure~\ref{fracphys} we show the difference in the species abundances between our nominal runs and the comparison runs including electron conduction and an increase in resolution. The comparisons, which are also presented in Table~\ref{fdata0}, show that neither electron conduction nor an increase in resolution has a substantial effect on our abundance results. 

\subsection{Parameter Dependencies}

Having explored  the evolution of three representative cases in detail, established convergence of abundances at a resolution at  $128^3,$ and shown that conduction has only a minor effect, we turn our attention to the full  suite of models. Table \ref{tab:simruns} gives the overall parameters of each of these runs, which span a large range of  columns, temperatures, and Mach numbers. In most cases, we found that heating from stirring and cooling from recombination are quickly balanced, as seen in the Med simulation presented in Fig.~\ref{evo}. However, if the stirring is sufficiently strong that the average temperature of the model exceeds $10^{5}$K these models will undergo thermal runaway. This can be explained by the fact that most elements have peaks in their cooling functions at  $\approx 10^{5}$K \citep[\eg][]{Gnat2012,Oppenheimer2013}. Therefore once the stirring passes this barrier, cooling can no longer balance turbulent heating and a meaningful steady state can no longer be achieved. 

A summary of our results is given in Table~\ref{fdata}, which shows the final abundances for all models that were able reach a steady state.   Inspection of these abundances reveals many interesting trends affecting commonly observed species. For example, without any contribution from photoionization, He$^+$ is often found at substantial levels, even when the average temperature is 10$^4$K, and in many of these cases the H$^+$ fraction can exceed 50\% (as seen when $\sigma_{1D}= 35$ km/s and $M_{\rm MW} = 2.8,$ or when $\sigma_{1D}=58$ km/s and  $M_{\rm MW} = 8.0$).  

On the other hand,  C$^{3+}$ mostly appears at substantial levels when the mean temperature is significantly higher than 10$^4$K, such as in the $\sigma_{1D}= 11.5$ km/s, $T_{\rm MW} = 7.0 \times 10^4$K, and $\sigma_{1D}= 20$ km/s,  $T_{\rm MW} = 5.5 \times 10^4$K runs.  Note however, that it is also seen in the high-mach number
$\sigma_{1D}= 58,$ km/s  run with $T_{\rm MW} = 1.1 \times 10^4$K, even though in CIE it is only expected when $T \geq 5 \times 10^4$K as shown in Fig.\ \ref{chemtest}.  Similarly, O$^{3+}$, which has a slightly lower ionization potential, is most abundant in runs with elevated temperatures or high mach numbers, although at even higher fractions than seen in C$^{3+}$. 

Interestingly, although its ionization potential is only 5.1 eV, substantial levels of neutral sodium are found in most of our simulations, particularly those with higher Mach numbers.  Inspections plots of the physical distribution of this ion show that it is mostly found in under-dense regions, which drop to relatively low temperatures through adiabatic cooling in expansions.  Finally, Mg$^+$  which recombines at 7.6 eV and is ionized at 15 eV is extremely abundant in all our simulations, with the exception of the highest temperature $\sigma_{1D}= 11.5$ km/s $T_{\rm MW} = 7.0 \times 10^4$K, and $\sigma_{1D}= 20$ km/s  $T_{\rm MW} = 5.5 \times 10^4$K cases.   In fact, in the absence of photoionization, the existence of significant, cospatial Mg$^{+}$, C$^{3+}$, and O$^{2+}$ is a clear indicator of a broad temperature-density PDF, caused by supersonic turbulence.

At highest Mach numbers, when the dispersions in temperature and density are large, regions in the simulation are able to reach conditions in which the fraction of molecular hydrogen may not be negligible.  Because parcels of gas move between regions with significantly different conditions on the timescale in which molecules are formed, calculating the H$_2$ fraction exactly would require solving a compete set of rate equations for H$_2$ formation within our simulations themselves. While this is beyond the scope of this work, we can nevertheless make a rough estimate of the molecular fractions calculating cell by cell abundances of H$^-,$ ${\rm H}_2^+$, and H$_2$ in the approximate case in which the local conditions last long enough for these additional species to come into equilibrium with the species tracked exactly by our simulations. In this case, in each cell we can compute
\begin{widetext}
\begin{minipage}{0.95\textwidth}
\ba
\label{eq:h2a}
F_{{\rm H}^-}  &=& \frac{k_1F_{\rm H}F_{\rm e^-}+k_{23}F_{{\rm H}_2^+}F_{\rm e^-} }{ (k_2+k_{15})F_{\rm H}+(k_5+k_{16})F_{{\rm H}^+}+k_{14}F_{\rm e^-}+(k_{21}+k_{22})F_{{\rm H}_2^+}+k_{28} F_{{\rm He}^+}+k_{29}F_{{\rm He}}} \\
\label{eq:h2b}
F_{{\rm H}_2^+}  &=& \frac{k_3F_{\rm H}F_{{\rm H}^+}+ k_7 F_{{\rm H}_2}F_{{\rm H}^+} + k_{16} F_{{\rm H}^+}F_{{\rm H}^-}  + k_{25}F_{{\rm H}_2}F_{{\rm He}^+}  }{k_4F_{\rm H}+k_6F_{\rm e^-}+ (k_{21} + k_{22})F_{{\rm H}^-}}\\
F_{{\rm H}_2} &=&  \frac{ k_2F_{{\rm H}^-}F_{\rm H} + k_4F_{\rm H}F_{{\rm H}_2^+} + k_{21}F_{{\rm H}^-}F_{{\rm H}_2^+} +k_RF_{\rm H}F_{\rm H}}
{ k_7F_{{\rm H}^+} + (k_8+k_{23})F_{\rm e^-} + k_9F_{\rm H}+k_{11} F_{\rm He} + (k_{24}+k_{25}) F_{{\rm He}^+}}
\label{eq:h2c}
\ea
\end{minipage}
\end{widetext}

where the temperature dependent reaction rates, $k_1$, $k_2,$ etc. are taken from the compilation presented in Appendix A of \cite{Glover2008}.   Furthermore, to estimate the formation of H$_2$ from  neutral-neutral surface reactions  on dust  we follow \cite{Glover2007}, eq.\ (36) which assumes that every collision between a metal atom and a grain results in a reaction. In that case, the reaction rate per unit volume for an atomic species i is given by $k_R = 3.0\times10^{-18}T^{1/2}/(1.0+4\times10^{-2}T^{1/2}+2\times10^{-3}T+8.0\times10^{-6}T^2)$. The resulting fractions are shown in Table  \ref{fdata}, denoted by asterisks because the are only approximate measurements. While the majority of hydrogen in our simulations remains atomic in all cases, these estimates indicate that in the highest Mach number runs, $\approx 1\%$ of the gas could be in the form of H$_2$.

\begin{table*}
\caption{Final steady-state abundances for our suite of models. Abundance values are given as $\log_{10}(F_i/F_T)$.
Here, ${\rm H^{-}}$,  ${\rm H_2^{+}}$, ${\rm {H_2}}$ are computed according to eqs.\ (\ref{eq:h2a}-\ref{eq:h2c}) and are thus approximate.}
\small
\begin{centering}
\tabcolsep=0.11cm
\scalebox{0.75}{
\begin{tabular}{l|rrrrrrrrrrrrrrr}
\hline
Col.\    (cm$^{-2}$)   & 1.1E15    & 1.1E16    & 1.1E17    & 3.0E16    & 1.1E17    & 3.0E17    & 1.5E17    & 7.6E17    & 1.5E18    & 1.5E18    & 4.6E18    & 1.5E19    & 1.1E19    & 3.0E19    & 1.1E20    \\
$\sigma_{1D}$  (km s$^{-1}$)  &  5.8      &  5.8      &  5.8      &  11.5     &  11.5     &  11.5     &  20.2     &  20.2     &  20.2     &  34.6     &  34.6     &  34.6     &  57.7     &  57.7     &  57.7     \\
$T_{DW}$  (10$^4$K) &  2.55  & 1.49  & 1.26  &  7.01 &  1.30 	&  1.14 &  5.54 &  1.01 &  0.87 	&  1.12 & 0.90 & 0.83 & 1.06 &  0.86 &  0.77 	\\
$M_{DW}$   & 0.29 & 0.51 & 0.61 & 0.44  & 1.06 & 1.26 & 0.91  &  2.39 & 2.78 & 4.13  &  4.63 & 5.32 & 7.98  & 8.79  & 10.6\\
\hline
\hline
${\rm H}$       & -1.03     & -0.23     & -0.03     & -3.96     & -0.19     & -0.10     & -3.47     & -0.17     & -0.09     & -0.42     & -0.19     & -0.09     & -0.37     & -0.15     & -0.07      \\
${\rm H^+}$     & -0.04     & -0.39     & -1.13     & 0.00      & -0.45     & -0.68     & 0.00      & -0.48     & -0.73     & -0.21     & -0.44     & -0.73     & -0.24     & -0.53     & -0.81      \\ 
${\rm He}$      & -0.00     & -0.00     & -0.00     & -2.87     & -0.17     & -0.01     & -2.54     & -0.00     & -0.00     & -0.01     & -0.00     & -0.00     & -0.03     & -0.00     & -0.00      \\
${\rm He^{+}}$  & -2.20     & -3.26     & -4.58     & -0.02     & -0.49     & -1.84     & -0.07     & -2.96     & -3.68     & -1.59     & -2.72     & -3.30     & -1.14     & -2.61     & -2.94      \\
${\rm He^{2+}}$ & -15.48    & -17.28    & -19.53    & -1.27     & -4.52     & -7.37     & -0.86     & -9.96     & -10.90    & -5.22     & -7.76     & -9.17     & -3.07     & -5.66     & -5.78      \\
${\rm C}$       & -1.51     & -0.67     & -0.21     & -3.71     & -0.67     & -0.46     & -3.10     & -0.50     & -0.28     & -0.63     & -0.34     & -0.19     & -0.37     & -0.19     & -0.12      \\
${\rm C^{+}}$   & -0.02     & -0.11     & -0.41     & -1.05     & -0.11     & -0.19     & -0.73     & -0.17     & -0.33     & -0.12     & -0.27     & -0.44     & -0.25     & -0.44     & -0.61      \\
${\rm C^{2+}}$  & -2.15     & -3.67     & -6.06     & -0.05     & -3.95     & -4.54     & -0.10     & -4.02     & -4.50     & -2.46     & -3.36     & -4.03     & -1.75     & -3.30     & -3.75      \\
${\rm C^{3+}}$  & -10.10    & -14.19    & -20.35    & -1.53     & -12.45    & -13.21    & -1.98     & -11.00    & -10.54    & -5.69     & -7.42     & -9.28     & -3.31     & -6.01     & -5.65      \\
${\rm C^{4+}}$  & -18.50    & -18.56    & -20.01    & -3.96     & -16.07    & -21.04    & -3.47     & -14.78    & -16.02    & -7.24     & -10.76    & -13.39    & -4.15     & -6.89     & -7.15      \\
${\rm O}$       & -1.25     & -0.28     & -0.05     & -3.49     & -0.50     & -0.25     & -3.07     & -0.26     & -0.10     & -0.53     & -0.20     & -0.09     & -0.37     & -0.15     & -0.08      \\
${\rm O^{+}}$   & -0.03     & -0.33     & -0.99     & -0.64     & -0.17     & -0.36     & -0.44     & -0.34     & -0.70     & -0.15     & -0.43     & -0.71     & -0.25     & -0.53     & -0.79      \\
${\rm O^{2+}}$  & -4.60     & -7.62     & -13.81    & -0.13     & -4.39     & -7.06     & -0.21     & -5.74     & -6.40     & -3.17     & -4.58     & -5.47     & -2.02     & -4.20     & -4.44      \\
${\rm O^{3+}}$  & -18.07    & -17.97    & -17.73    & -1.54     & -14.15    & -15.07    & -1.88     & -11.64    & -11.91    & -5.62     & -8.00     & -10.23    & -3.20     & -5.82     & -5.70      \\
${\rm O^{4+}}$  & -19.46    & -18.92    & -18.46    & -5.51     & -24.67    & -25.59    & -5.92     & -20.29    & -19.12    & -8.90     & -12.32    & -16.24    & -4.73     & -7.68     & -7.61      \\
${\rm O^{5+}}$  & -19.54    & -18.72    & -18.22    & -12.31    & -26.39    & -26.34    & -11.67    & -23.90    & -26.03    & -11.26    & -16.89    & -21.61    & -6.21     & -9.13     & -10.90     \\
${\rm O^{6+}}$  & -19.51    & -18.85    & -18.54    & -18.62    & -26.33    & -26.33    & -13.14    & -26.34    & -26.34    & -14.61    & -22.27    & -26.33    & -7.53     & -10.96    & -14.93     \\
${\rm O^{7+}}$  & -21.43    & -19.68    & -19.64    & -20.34    & -26.54    & -26.54    & -20.89    & -26.54    & -26.55    & -26.55    & -26.56    & -26.57    & -15.17    & -19.93    & -23.88     \\
${\rm Na}$      & -5.31     & -4.26     & -3.87     & -5.76     & -3.88     & -3.51     & -5.86     & -2.91     & -2.45     & -2.75     & -2.19     & -1.98     & -2.10     & -1.86     & -1.57      \\
${\rm Na^{+}}$  & -0.00     & -0.00     & -0.00     & -0.00     & -0.00     & -0.00     & -0.00     & -0.00     & -0.00     & -0.00     & -0.00     & -0.00     & -0.00     & -0.01     & -0.01      \\
${\rm Mg}$      & -2.97     & -1.40     & -0.98     & -7.23     & -1.29     & -1.04     & -6.37     & -0.88     & -0.67     & -0.98     & -0.61     & -0.48     & -0.65     & -0.44     & -0.35      \\
${\rm Mg^{+}}$  & -0.83     & -0.18     & -0.06     & -3.84     & -0.16     & -0.09     & -3.41     & -0.13     & -0.12     & -0.25     & -0.18     & -0.20     & -0.31     & -0.25     & -0.29      \\
${\rm Mg^{2+}}$ & -0.07     & -0.53     & -1.57     & -0.00     & -0.60     & -1.03     & -0.00     & -0.93     & -1.42     & -0.48     & -1.00     & -1.38     & -0.55     & -1.13     & -1.46      \\   
${\rm H^{-*}}$  & -8.39  	& -7.42  	& -7.39 	& -11.79  	& -7.36  	& -7.32 	& -11.13  	& -7.38  	& -7.36  	& -7.61  	& -7.42  	& -7.37  	& -7.58  	  & -7.40  	& -7.40      \\
${\rm H_2^{+*}}$& -7.08  	& -6.90  	& -7.51  	& -9.38  	& -6.99  	& -7.14  	& -9.07  	& -7.08  	& -7.20  	& -7.10  	& -7.11  	& -7.22  	& -7.18  	  & -7.16  	& -7.30      \\
${\rm {H_2}^*}$ & -7.85  	& -6.24  	& -6.15 	& -14.08  	& -5.84  	& -3.90 	& -12.84  	& -3.16  	& -2.14  	& -2.86  	& -2.17  	& -1.84  	& -2.21     & -1.86  	& -1.60      
\label{fdata}
\end{tabular}
}
\end{centering}
\end{table*}

In addition to Table~\ref{fdata}  we make available online the data files from each run\footnote{http://zofia.sese.asu.edu/\~{}evan/turbspecies/}. Each file presents a variety of information, including the true abundances and the simple abundance estimates presented in \S\ref{IS}. We also give Doppler parameters for each species, $i$, defined as,
\be
b_i^2 = \sigma_{i,{\rm 1D}}^2 + 2k_bT_{\rm MW}/A_{i}m_{\rm H},
\ee
where $\sigma_{\rm i,1D}$ is the 1D velocity dispersion, $k_b$ is the Boltzmann's constant, and $A_i$ is the ion atomic mass. 
While the masses of hydrogen and helium are small enough that the temperature term makes a substantial contribution,  the Doppler parameters of the heavier elements are very close to $\sigma_{i,\rm{1D}}$ in general,
providing a good measurement of the local velocity dispersion.

\section{Conclusions}

Turbulence pervades the Universe, often moving at supersonic speeds due to the high efficiency of radiative cooling. These random motions provide overall support against gravity, but also concentrate a portion of the material to very high densities, giving rise to a multiphase distribution with unique thermodynamic properties.  This can have a dramatic effect on the chemical makeup of the medium. Specifically, if the recombination time for a given species is long compared to the eddy turnover time, it cannot reach an equilibrium state before it is further acted upon by the turbulence. This creates a situation in which the final ionization state is not only a function of the temperature and density, but also a function of the rate at which parcels of gas move through these conditions.

To study how these abundances are altered by turbulence, we have implemented a nonequilibrium atomic chemistry package within FLASH. This package tracks the evolution  of six elements and 24 separate ionization states. In addition, we have used the method of \cite{Gnat2012} to derive ion-by-ion cooling curves for each of the ions under consideration that allows us to follow the thermodynamic evolution of the gas. The result is a very fast and efficient package, which we are able to run for many eddy turnover times for many cases.

Using this tool,
we have performed a suite of direct numerical simulations of solenoidally-driven turbulence over a range from 1D velocity dispersion of $\sigma_{\rm 1D}$= 6 to 58 km s$^{-1}$ and the product of the mean density and turbulent length scale from 10$^{16}$ to 10$^{20}$ cm$^{-2}$ for solar metallicity gas, concentrating on three representative models. As found by isothermal models of driven turbulence, the gas approximates a lognormal distribution, whose logarithmic density variance in the supersonic case is well approximated by $\sigma_{s}^2 = {\rm ln}( 1 + b^2 M^2).$  On the other hand, this expression overestimates the variance at subsonic Mach numbers, and the $b=0.53$ value that best fits our data is significantly different from the $b=1/3$ value measured in solenoidally-driven, isothermal turbulence. 

We compare the final steady state abundances in our simulations to those obtained assuming the gas is in collisional ionization equilibrium, using both the mean temperature and the full temperature PDF. We find that at low mach numbers the estimates agree to within a factor of $\approx 2$ for most species, save for He$^{2+}$ and \ctwo, which show large deviations due to their long recombination times. At intermediate mach numbers, several species such as He, C, C$^{+}$, C$^{3+}$, O$^{3+}$, and Mg, differ by a factor of $\approx 2$ from our simple estimates, while He$^{2+}$  and C$^{4+}$ are $\approx 5$ more abundant than our simples estimates suggest. Finally, for very high mach numbers, the abundances can vary by many orders of magnitude from simple estimates.  Neither increasing the resolution by a factor of two or including of thermal electron conduction has a significant effect on these abundances.

These results underscore the fact that transsonic and supersonic turbulence can drastically alter the abundances and that only nonequilibrium calculations can predict these changes accurately. Thus we make make all of the derived properties from our  models available online. In particular, we present the logarithmic density statistics and other hydrodynamic  quantities, such as Mach number and average temperature.  We also give the final abundances for each species, the abundance values from the two simple estimates, and species by species Doppler parameters. 

In future work we, plan on increasing our network to include ions of nitrogen (N-N$^{5+}$), silicon (Si-Si$^{4+}$), and iron (Fe-Fe$^{3+}$) and their associated cooling as well as the effects of photoionization.  Using this increased network, we will run similar models as those presented here, again compiling important statistical properties and the final ionization structure of the gas. This will result in a large set of tables, useful for a variety of theoretical and observational applications.

\acknowledgements

We would like to thank Christopher Matzner, Cody Raskin, Eve Ostriker, Robert J. Thacker, and David Williamson for helpful discussions. ES gratefully acknowledges Joanne Cohn, Eliott Quatert, and the UC Berkeley Theoretical Astronomy Center, Uros Seljak and the Lawrence Berkeley National Lab Cosmology group, and the organizers of the GravityÕs Loyal Opposition: The Physics of Star Formation Feedback at the Kavli Institute for Theoretical Physics, for hosting him during the period when much of this work was carried out. The software used in this work was in part developed by the DOE NNSA-ASC OASCR Flash Center at the University of Chicago. This work was performed under the auspices of the U.S. Department of Energy by Lawrence Livermore National Laboratory under Contract DE-AC52-07NA27344. LLNL-JRNL-662155. ES was supported by NSF grant AST11-03608, and PHY11-25915. The figures and analysis presented here were created using the {\bf yt} analysis package \citep{Turk2011}. 

\bibliographystyle{apj}
\bibliography{ms.bib}

\begin{thebibliography}{}
\expandafter\ifx\csname natexlab\endcsname\relax\def\natexlab#1{#1}\fi

\bibitem[{{Allen} {et~al.}(2008){Allen}, {Groves}, {Dopita}, {Sutherland}, \&
  {Kewley}}]{Allen2008}
{Allen}, M.~G., {Groves}, B.~A., {Dopita}, M.~A., {Sutherland}, R.~S., \&
  {Kewley}, L.~J. 2008, \apjs, 178, 20

\bibitem[{{Altun} {et~al.}(2004){Altun}, {Yumak}, {Badnell}, {Colgan}, \&
  {Pindzola}}]{Altun2004}
{Altun}, Z., {Yumak}, A., {Badnell}, N.~R., {Colgan}, J., \& {Pindzola}, M.~S.
  2004, \aap, 420, 775

\bibitem[{{Altun} {et~al.}(2006){Altun}, {Yumak}, {Badnell}, {Loch}, \&
  {Pindzola}}]{Altun2006}
{Altun}, Z., {Yumak}, A., {Badnell}, N.~R., {Loch}, S.~D., \& {Pindzola}, M.~S.
  2006, \aap, 447, 1165

\bibitem[{Bader \& Deuflhard(1983)}]{Bader1983}
Bader, G., \& Deuflhard, P. 1983, Numerische Mathematik, 41, 373

\bibitem[{{Badnell}(2006{\natexlab{a}})}]{Badnell2006H}
{Badnell}, N.~R. 2006{\natexlab{a}}, \aap, 447, 389

\bibitem[{{Badnell}(2006{\natexlab{b}})}]{Badnell2006RR}
---. 2006{\natexlab{b}}, \apjs, 167, 334

\bibitem[{{Bautista} \& {Badnell}(2007)}]{Bautista2007}
{Bautista}, M.~A., \& {Badnell}, N.~R. 2007, \aap, 466, 755

\bibitem[{{Bovino} {et~al.}(2013){Bovino}, {Grassi}, {Latif}, \&
  {Schleicher}}]{Bovino2013}
{Bovino}, S., {Grassi}, T., {Latif}, M.~A., \& {Schleicher}, D.~R.~G. 2013,
  \mnras, 434, L36

\bibitem[{{Braginskii}(1958)}]{Braginskii1958}
{Braginskii}, S.~I. 1958, Soviet Journal of Experimental and Theoretical
  Physics, 6, 358

\bibitem[{{Colella} \& {Glaz}(1985)}]{Colella1985}
{Colella}, P., \& {Glaz}, H.~M. 1985, Journal of Computational Physics, 59, 264

\bibitem[{{Colgan} {et~al.}(2004){Colgan}, {Pindzola}, \&
  {Badnell}}]{Colgan2004}
{Colgan}, J., {Pindzola}, M.~S., \& {Badnell}, N.~R. 2004, \aap, 417, 1183

\bibitem[{{Colgan} {et~al.}(2003){Colgan}, {Pindzola}, {Whiteford}, \&
  {Badnell}}]{Colgan2003}
{Colgan}, J., {Pindzola}, M.~S., {Whiteford}, A.~D., \& {Badnell}, N.~R. 2003,
  \aap, 412, 597

\bibitem[{{Collins} {et~al.}(2011){Collins}, {Padoan}, {Norman}, \&
  {Xu}}]{Collins2011}
{Collins}, D.~C., {Padoan}, P., {Norman}, M.~L., \& {Xu}, H. 2011, \apj, 731,
  59

\bibitem[{{Cowie} \& {McKee}(1977)}]{Cowie1977}
{Cowie}, L.~L., \& {McKee}, C.~F. 1977, \apj, 211, 135

\bibitem[{Duff {et~al.}(1986)Duff, Erisman, \& Reid}]{Duff1986}
Duff, I.~S., Erisman, A.~M., \& Reid, J.~K. 1986, Direct methods for sparse
  matrices (Clarendon Press Oxford)

\bibitem[{{Einfeldt} {et~al.}(1991){Einfeldt}, {Roe}, {Munz}, \&
  {Sjogreen}}]{Einfeldt1991}
{Einfeldt}, B., {Roe}, P.~L., {Munz}, C.~D., \& {Sjogreen}, B. 1991, Journal of
  Computational Physics, 92, 273

\bibitem[{{Falkovich}(1994)}]{Falkovich1994}
{Falkovich}, G. 1994, Physics of Fluids, 6, 1411

\bibitem[{{Federrath} {et~al.}(2008){Federrath}, {Klessen}, \&
  {Schmidt}}]{Federrath2008}
{Federrath}, C., {Klessen}, R.~S., \& {Schmidt}, W. 2008, \apjl, 688, L79

\bibitem[{{Federrath} {et~al.}(2010){Federrath}, {Roman-Duval}, {Klessen},
  {Schmidt}, \& {Mac Low}}]{Federrath2010}
{Federrath}, C., {Roman-Duval}, J., {Klessen}, R.~S., {Schmidt}, W., \& {Mac
  Low}, M.-M. 2010, \aap, 512, A81

\bibitem[{{Ferland} {et~al.}(1998){Ferland}, {Korista}, {Verner}, {Ferguson},
  {Kingdon}, \& {Verner}}]{Ferland1998}
{Ferland}, G.~J., {Korista}, K.~T., {Verner}, D.~A., {et~al.} 1998, \pasp, 110,
  761

\bibitem[{{Fryxell} {et~al.}(2000){Fryxell}, {Olson}, {Ricker}, {Timmes},
  {Zingale}, {Lamb}, {MacNeice}, {Rosner}, {Truran}, \& {Tufo}}]{Fryxell2000}
{Fryxell}, B., {Olson}, K., {Ricker}, P., {et~al.} 2000, \apjs, 131, 273

\bibitem[{{Glover} \& {Abel}(2008)}]{Glover2008}
{Glover}, S.~C.~O., \& {Abel}, T. 2008, \mnras, 388, 1627

\bibitem[{{Glover} {et~al.}(2010){Glover}, {Federrath}, {Mac Low}, \&
  {Klessen}}]{Glover2010}
{Glover}, S.~C.~O., {Federrath}, C., {Mac Low}, M.-M., \& {Klessen}, R.~S.
  2010, \mnras, 404, 2

\bibitem[{{Glover} \& {Jappsen}(2007)}]{Glover2007}
{Glover}, S.~C.~O., \& {Jappsen}, A.-K. 2007, \apj, 666, 1

\bibitem[{{Gnat} \& {Ferland}(2012)}]{Gnat2012}
{Gnat}, O., \& {Ferland}, G.~J. 2012, \apjs, 199, 20

\bibitem[{{Gray} \& {Scannapieco}(2010)}]{Gray2010}
{Gray}, W.~J., \& {Scannapieco}, E. 2010, \apj, 718, 417

\bibitem[{{Gray} \& {Scannapieco}(2013)}]{Gray2013}
---. 2013, \apj, 768, 174

\bibitem[{{Green} {et~al.}(2012){Green}, {Froning}, {Osterman}, \& {\it et al.\
  }}]{Green2012}
{Green}, J.~C., {Froning}, C.~S., {Osterman}, S., \& {\it et al.\ }. 2012,
  \apj, 744, 60

\bibitem[{{Ishihara} {et~al.}(2009){Ishihara}, {Gotoh}, \&
  {Kaneda}}]{Ishihara2009}
{Ishihara}, T., {Gotoh}, T., \& {Kaneda}, Y. 2009, Annual Review of Fluid
  Mechanics, 41, 165

\bibitem[{{Kewley} {et~al.}(2006){Kewley}, {Groves}, {Kauffmann}, \&
  {Heckman}}]{Kewley2006}
{Kewley}, L.~J., {Groves}, B., {Kauffmann}, G., \& {Heckman}, T. 2006, \mnras,
  372, 961

\bibitem[{{Klessen}(2000)}]{Klessen2000}
{Klessen}, R.~S. 2000, \apj, 535, 869

\bibitem[{{Kritsuk} {et~al.}(2007){Kritsuk}, {Norman}, {Padoan}, \&
  {Wagner}}]{Kritsuk2007}
{Kritsuk}, A.~G., {Norman}, M.~L., {Padoan}, P., \& {Wagner}, R. 2007, \apj,
  665, 416

\bibitem[{{Lemaster} \& {Stone}(2008)}]{Lemaster2008}
{Lemaster}, M.~N., \& {Stone}, J.~M. 2008, \apjl, 682, L97

\bibitem[{{Li} {et~al.}(2003){Li}, {Klessen}, \& {Mac Low}}]{Yi2003}
{Li}, Y., {Klessen}, R.~S., \& {Mac Low}, M.-M. 2003, \apj, 592, 975

\bibitem[{{Mac Low} \& {Klessen}(2004)}]{MacLow2004}
{Mac Low}, M.-M., \& {Klessen}, R.~S. 2004, Reviews of Modern Physics, 76, 125

\bibitem[{{Mitnik} \& {Badnell}(2004)}]{Mitnik2004}
{Mitnik}, D.~M., \& {Badnell}, N.~R. 2004, \aap, 425, 1153

\bibitem[{{Moin} \& {Mahesh}(1998)}]{Moin1998}
{Moin}, P., \& {Mahesh}, K. 1998, Annual Review of Fluid Mechanics, 30, 539

\bibitem[{{Molina} {et~al.}(2012){Molina}, {Glover}, {Federrath}, \&
  {Klessen}}]{Molina2012}
{Molina}, F.~Z., {Glover}, S.~C.~O., {Federrath}, C., \& {Klessen}, R.~S. 2012,
  \mnras, 423, 2680

\bibitem[{{Oppenheimer} \& {Schaye}(2013)}]{Oppenheimer2013}
{Oppenheimer}, B.~D., \& {Schaye}, J. 2013, \mnras, 434, 1043

\bibitem[{{Orszag} \& {Kells}(1980)}]{Orszag1980}
{Orszag}, S.~A., \& {Kells}, L.~C. 1980, Journal of Fluid Mechanics, 96, 159

\bibitem[{{Osterbrock}(1989)}]{Osterbrock1989}
{Osterbrock}, D.~E. 1989, {Astrophysics of gaseous nebulae and active galactic
  nuclei}

\bibitem[{{Osterbrock} \& {Ferland}(2006)}]{Osterbrock2006}
{Osterbrock}, D.~E., \& {Ferland}, G.~J. 2006, {Astrophysics of gaseous nebulae
  and active galactic nuclei}

\bibitem[{{Ostriker} {et~al.}(2001){Ostriker}, {Stone}, \&
  {Gammie}}]{Ostriker2001}
{Ostriker}, E.~C., {Stone}, J.~M., \& {Gammie}, C.~F. 2001, \apj, 546, 980

\bibitem[{{Padoan} \& {Nordlund}(2011)}]{Padoan2011}
{Padoan}, P., \& {Nordlund}, {\AA}. 2011, \apj, 730, 40

\bibitem[{{Padoan} {et~al.}(1997){Padoan}, {Nordlund}, \& {Jones}}]{Padoan1997}
{Padoan}, P., {Nordlund}, A., \& {Jones}, B.~J.~T. 1997, \mnras, 288, 145

\bibitem[{{Pan} \& {Scannapieco}(2010)}]{Pan2010}
{Pan}, L., \& {Scannapieco}, E. 2010, \apj, 721, 1765

\bibitem[{{Peeples} {et~al.}(2014){Peeples}, {Werk}, {Tumlinson},
  {Oppenheimer}, {Prochaska}, {Katz}, \& {Weinberg}}]{Peeples2014}
{Peeples}, M.~S., {Werk}, J.~K., {Tumlinson}, J., {et~al.} 2014, \apj, 786, 54

\bibitem[{{Press} {et~al.}(1992){Press}, {Teukolsky}, {Vetterling}, \&
  {Flannery}}]{Press1992}
{Press}, W.~H., {Teukolsky}, S.~A., {Vetterling}, W.~T., \& {Flannery}, B.~P.
  1992, {Numerical recipes in FORTRAN. The art of scientific computing}

\bibitem[{{Price} {et~al.}(2011){Price}, {Federrath}, \& {Brunt}}]{Price2011}
{Price}, D.~J., {Federrath}, C., \& {Brunt}, C.~M. 2011, \apjl, 727, L21

\bibitem[{{Saury} {et~al.}(2014){Saury}, {Miville-Desch{\^e}nes}, {Hennebelle},
  {Audit}, \& {Schmidt}}]{Saury2014}
{Saury}, E., {Miville-Desch{\^e}nes}, M.-A., {Hennebelle}, P., {Audit}, E., \&
  {Schmidt}, W. 2014, \aap, 567, A16

\bibitem[{{Scannapieco} {et~al.}(2012){Scannapieco}, {Gray}, \&
  {Pan}}]{Scannapieco2012}
{Scannapieco}, E., {Gray}, W.~J., \& {Pan}, L. 2012, \apj, 746, 57

\bibitem[{{Schmidt} {et~al.}(2009){Schmidt}, {Federrath}, {Hupp}, {Kern}, \&
  {Niemeyer}}]{Schmidt2009}
{Schmidt}, W., {Federrath}, C., {Hupp}, M., {Kern}, S., \& {Niemeyer}, J.~C.
  2009, \aap, 494, 127

\bibitem[{{Soto} \& {Martin}(2012)}]{Soto2012}
{Soto}, K.~T., \& {Martin}, C.~L. 2012, \apjs, 203, 3

\bibitem[{{Spitzer}(1962)}]{Spitzer1962}
{Spitzer}, L. 1962, {Physics of Fully Ionized Gases}

\bibitem[{{Sur} {et~al.}(2014){Sur}, {Pan}, \& {Scannapieco}}]{Sur2014}
{Sur}, S., {Pan}, L., \& {Scannapieco}, E. 2014, \apj, 784, 94

\bibitem[{Toro {et~al.}(1994)Toro, Spruce, \& Speares}]{Toro1994}
Toro, E., Spruce, M., \& Speares, W. 1994, Shock Waves, 4, 25

\bibitem[{Toro(1999)}]{Toro1999}
Toro, E.~F. 1999, Riemann solvers and numerical methods for fluid dynamics,
  Vol.~16 (Springer)

\bibitem[{{Tumlinson} {et~al.}(2013){Tumlinson}, {Thom}, {Werk}, \& {{\it et
  al.\ }}}]{Tumlinson2013}
{Tumlinson}, J., {Thom}, C., {Werk}, J.~K., \& {{\it et al.\ }}. 2013, \apj,
  777, 59

\bibitem[{{Turk} {et~al.}(2011){Turk}, {Smith}, {Oishi}, {Skory}, {Skillman},
  {Abel}, \& {Norman}}]{Turk2011}
{Turk}, M.~J., {Smith}, B.~D., {Oishi}, J.~S., {et~al.} 2011, \apjs, 192, 9

\bibitem[{Uhlenbeck \& Ornstein(1930)}]{Uhlenbeck1930}
Uhlenbeck, G.~E., \& Ornstein, L.~S. 1930, Phys. Rev., 36, 823

\bibitem[{{Vazquez-Semadeni}(1994)}]{Vazquez1994}
{Vazquez-Semadeni}, E. 1994, \apj, 423, 681

\bibitem[{{Verner} \& {Ferland}(1996)}]{Verner1996}
{Verner}, D.~A., \& {Ferland}, G.~J. 1996, \apjs, 103, 467

\bibitem[{{Vincent} \& {Meneguzzi}(1991)}]{Vincent1991}
{Vincent}, A., \& {Meneguzzi}, M. 1991, in Lecture Notes in Physics, Berlin
  Springer Verlag, Vol. 380, IAU Colloq. 130: The Sun and Cool Stars. Activity,
  Magnetism, Dynamos, ed. I.~{Tuominen}, D.~{Moss}, \& G.~{R{\"u}diger}, 75

\bibitem[{{Voronov}(1997)}]{Voronov1997}
{Voronov}, G.~S. 1997, Atomic Data and Nuclear Data Tables, 65, 1

\bibitem[{{Walch} {et~al.}(2011){Walch}, {W{\"u}nsch}, {Burkert}, {Glover}, \&
  {Whitworth}}]{Walch2011}
{Walch}, S., {W{\"u}nsch}, R., {Burkert}, A., {Glover}, S., \& {Whitworth}, A.
  2011, \apj, 733, 47

\bibitem[{{Walch} {et~al.}(2014){Walch}, {Girichidis}, {Naab}, {Gatto},
  {Glover}, {W{\"u}nsch}, {Klessen}, {Clark}, {Peters}, \&
  {Baczynski}}]{Walch2014}
{Walch}, S.~K., {Girichidis}, P., {Naab}, T., {et~al.} 2014, ArXiv e-prints,
  arXiv:1412.2749

\bibitem[{{Werk} {et~al.}(2013){Werk}, {Prochaska}, {Thom}, \& {\it et
  al.}}]{Werk2013}
{Werk}, J.~K., {Prochaska}, J.~X., {Thom}, C., \& {\it et al.} 2013, \apjs,
  204, 17

\bibitem[{{Werk} {et~al.}(2014){Werk}, {Prochaska}, {Tumlinson}, \& {\it et
  al.\ }}]{Werk2014}
{Werk}, J.~K., {Prochaska}, J.~X., {Tumlinson}, J., \& {\it et al.\ }. 2014,
  \apj, 792, 8

\bibitem[{{Wolfire} {et~al.}(2003){Wolfire}, {McKee}, {Hollenbach}, \&
  {Tielens}}]{Wolfire2003}
{Wolfire}, M.~G., {McKee}, C.~F., {Hollenbach}, D., \& {Tielens}, A.~G.~G.~M.
  2003, \apj, 587, 278

\bibitem[{{Zatsarinny} {et~al.}(2004{\natexlab{a}}){Zatsarinny}, {Gorczyca},
  {Korista}, {Badnell}, \& {Savin}}]{Zat2004b}
{Zatsarinny}, O., {Gorczyca}, T.~W., {Korista}, K., {Badnell}, N.~R., \&
  {Savin}, D.~W. 2004{\natexlab{a}}, \aap, 426, 699

\bibitem[{{Zatsarinny} {et~al.}(2004{\natexlab{b}}){Zatsarinny}, {Gorczyca},
  {Korista}, {Badnell}, \& {Savin}}]{Zat2004a}
{Zatsarinny}, O., {Gorczyca}, T.~W., {Korista}, K.~T., {Badnell}, N.~R., \&
  {Savin}, D.~W. 2004{\natexlab{b}}, \aap, 417, 1173

\end{thebibliography}

\end{document}